\documentclass[preprint]{aastex}

\slugcomment{To appear in the Astrophysical Journal}

\shorttitle{Poisson equation solver}
\shortauthors{Matsumoto and Hanawa}

\begin{document}

\title{A Fast Algorithm for Solving the Poisson Equation on a
Nested Grid}

\author{Tomoaki Matsumoto}
\affil{Department of Humanity and Environment, Hosei University,
Fujimi, Chiyoda-ku, Tokyo 102-8160, Japan}
\email{matsu@i.hosei.ac.jp}

\and

\author{Tomoyuki Hanawa}
\affil{Department of Astrophysics, School of Science, Nagoya
University, Chikusa-ku, Nagoya 464-8602, Japan}
\email{hanawa@a.phys.nagoya-u.ac.jp}

\begin{abstract}
We present a numerical method for solving the Poisson
equation on a nested grid.  The nested grid consists of
uniform grids having different grid spacing and is designed
to cover the space closer to the center with a finer grid.
Thus our numerical method is suitable for computing the
gravity of a centrally condensed object.
It consists of two parts: the difference scheme for the
Poisson equation on the nested grid and 
the multi-grid iteration algorithm.
It has three advantages: accuracy, fast convergence, and 
scalability.  First it computes the gravitational potential
of a close binary accurately up to the quadraple moment, 
even when the binary is resolved only in the fine grids.
Second residual decreases by a factor of 300 or more
by each iteration.  
We confirmed experimentally that 
the iteration converges always to the exact solution of 
the difference equation.  Third the computation load of the iteration
is proportional to the total number of the cells in the
nested grid.  Thus our method gives a good solution at the
minimum expense when the nested grid is large.
The difference scheme is applicable also to the adaptive
mesh refinement in which cells of different sizes are used
to cover a domain of computation.
\end{abstract}


\section{Introduction}

Astronomical objects such as stars, clouds, and galaxies
have enormous dynamic range both in density and 
in size.  To illustrate this enormous dynamic range,
we consider star formation as an example.  Stars form
in molecular clouds of which the mean density is 
10$^3$ atoms~cm$^{-3}$.  The molecular clouds contain
condensations named molecular cloud cores, from which
stars form owing to the self-gravity.  The molecular
cloud cores have typical size of 10$^{17}$ cm and typical
density of 10$^5$ atoms~cm$^{-3}$. On the other hand,
the central density of a star is 10$^{11}$ atoms cm$^{-3}$
at the very beginning of its protostellar stage and
the present Sun has the central density of 10$^{26}$ 
atoms~cm$^{-3}$.  The radius of a protostar is 
10$^{14}$ cm when the central density is 10$^{12}$ 
atoms~cm$^{-3}$.  As the central density
increases, the radius decreases down to 10$^{10}$
-- 10$^{12}$ cm until the star reaches its main sequence
(hydrogen burning) stage. 

This enormous dynamic range restrict us to achieve 
high spatial resolution only in the small regions 
of interest.  To generate finer grids 
in the region of interest, people have developed
various mesh generation methods. 
Adaptive mesh
refinement (AMR) and nested grids (NG) are typical of the
mesh generation methods developed in the past decade.
AMR and NG generate finer grids hierarchically in 
the region of interest.  AMR was invented by \citet{berger84}
and has been advanced by many researchers.
NG is a variant of AMR and
generates only one sub-grid at each hierarchical level,
while AMR has no restriction on the number of sub-grids.
AMR and NG have succeeded in simulations of star formation
and galaxy formation in which compact objects form by
condensation of diffuse clouds.

Some recent numerical simulations on star formation and
cosmology apply either AMR or NG to achieve high spatial
resolution.  \citet{truelove97} studied gravitational
collapse of a molecular cloud core with AMR to
resolve fragmentation of the highly condensed cloud core.
Since then AMR is frequently used in numerical simulations
of fragmentation during gravitational collapse
\citep{truelove98,boss00}.
Using NG \citet{burkert93,burkert96} studied fragmentation
of a centrally condensed protostar.  They succeeded in 
resolving spiral arms formed by the self-gravitational
instability in the protoplanetary disk.   
Using NG \citet{tomisaka98} computed 
gravitational collapse of rotating magnetized gas cloud
and found magnetohydrodynamical driven outflow emanating
from a very compact central disk.  
\citet{norman99} has reviewed application
of AMR to cosmological simulations. 

Though simulations based on AMR and NG are successful,
some technical problems still remain for AMR and NG.
One of them is a numerical algorithm for solving the Poisson
equation on a grid in which a cell faces several smaller cells.
In other words, difficulty arises at boundaries between the 
regions covered with different size cells.  
For later convenience we name these boundaries the
grid level boundaries. In most AMR and NG of three dimensions,
a parent cell faces four child cells at each grid level
boundary in case of three dimensions. 

When cells are uniform on a grid, a simple central difference 
scheme gives us second order accuracy and the difference equation
can be solved fast with the multi-grid iteration or some
other methods.  Since the simple central difference 
breaks down at the grid level boundary, we need some modifications
at the boundaries.  An ideal scheme should provide an accurate 
solution and require minimum computation load.  
As shown later, some schemes published in recent literature
lack in accuracy.

One might think that we could obtain a good solution by relatively
small computation cost if we would solve the Poisson equation
successively from coarse grids to fine grids.
One can solve the Poisson equation on the coarsest grid 
if an appropriate boundary is given at the boundary of
the grid.  Then the solution gives boundary conditions for
the sub-grid if we interpolate it appropriately.  
Thus we can solve the Poisson equation on the sub-grid and
repeat the same procedure down to the finest grid.
This algorithm requires rather small computation load but
the solution is not accurate enough.

To illustrate the problem of the algorithm mentioned above,
we consider binary or multiple stars of which separation
is too short to resolve in a coarse grid.  For simplicity we
assume that the binary or multiple stars dominate the
gravity.  At a large distance from them, the gravity is
approximately the sum of the point mass gravity and the quadraple
moment of gravity.  If we compute the gravity with a coarse
grid without any knowledge on the fine structure, we will
miss the quadraple moment of gravity.   Although it is smaller
than the point mass gravity, the quadraple moment of gravity
is the leading term in the gravitational torque.  We need
to take account of mass distribution in a fine grid to 
evaluate the gravitational torque.

In this paper we present a difference equation for
the Poisson equation on a nested grid.   The solution of
this difference equation is accurate enough in the sense
that it reproduces the quadraple moment of gravity
quantitatively even in a coarse grid.  Moreover the
difference equation can be solved fast with the multi-grid
iteration.   The computation load is scalable, i.e., 
proportional to the total number of cells involved in
the nested grid.  

The above difference equation is also applicable to AMR.
The difference equation is likely
to be solved fast also with a multi-grid iteration.

In \S 2 we describe our difference equation.  In \S 3 we
denote our algorithm for solving the difference equation.
In \S 4 we show the performance of our scheme with emphasis
on the accuracy of the solution and speed of computation.
In \S 5 we discuss reason for success of our difference equation
in reproducing the quadraple moment of gravity.  We also
discuss extension of our method to AMR.

\section{Difference Equation}

\begin{figure}
\epsscale{0.5}
\plotone{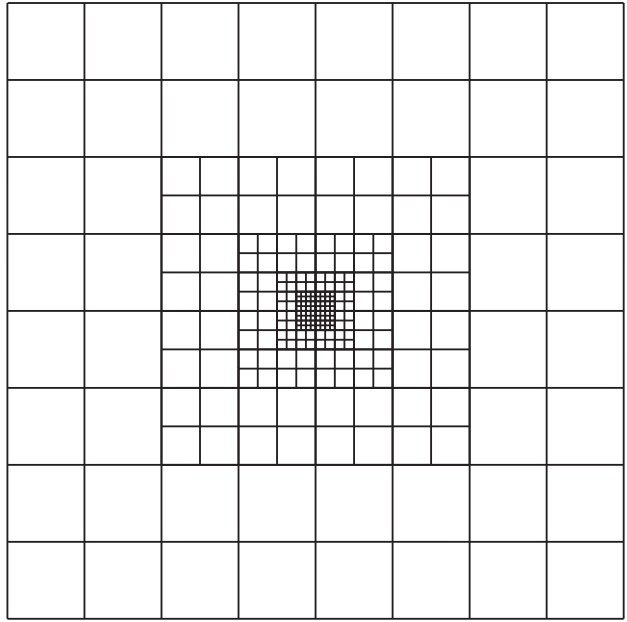}
\caption{The cross section of our nested grid is shown. 
The lines denote the cell boundaries.  The region closer to
the center is covered with a smaller cells.
This nested grid has the parameters, $ N $ = 8 and 
$ \ell _{\rm max} $ = 5.}
\label{nestg.eps}
\end{figure}

First we introduce our nested grid and define the coordinates.
The nested grid consists of uniform grids each of which
has cubic cells.  The center of each cell is located at
\begin{eqnarray}
x _i ^{(\ell)} & = & 2 ^{-\ell+1} \, (i \, + \, 1/2) \, h \, , \\
y _j ^{(\ell)} & = & 2 ^{-\ell+1} \, (j \, + \, 1/2) \, h \, , \\
z _k ^{(\ell)} & = & 2 ^{-\ell+1} \, (k \, + \, 1/2) \, h \, ,
\end{eqnarray}
where $ i $, $ j $, $ k $ and $ \ell $ are integer and labels.
The symbol, $ h $, denotes the grid spacing in the coarsest
grid at the level of $ \ell \, = \, 1 $.  As $ \ell $ increases
by one, the grid spacing reduces to half. 
We define the range of $ i $, $ j $, $ k $, and $ \ell $ as
\begin{eqnarray}
- N/2 \; \le & i & \le \; N/2 \, - \, 1 \; , \\
- N/2 \; \le & j & \le \; N/2 \, - \, 1 \; , \\
- N/2 \; \le & k & \le \; N/2 \; - \, 1 , \\
1 \; \le & \ell & \le \; \ell _{\rm max} \; , \\
\end{eqnarray}
in this paper.   Thus our nested grid has 
$ \ell _{\rm max} \, \times  \, N ^3 $ cells in total. 
To use the multi-grid iteration effectively, we take
$ N \, = \, 2 ^n $ in our computation.
The coarsest grid
covers the volume of $ L ^3 $, where
$ L \, = \, N  \, h $.  The finest grid has grid spacing
of $ 2 ^{\ell _{\rm max} \, - \, 1} \, h $.
Since all these grids are concentric, central regions are covered
multiply with these grids. 
When a volume is covered with different size cells, we adopt
a quantity in the smallest cell in the final solution.
Figure \ref{nestg.eps} illustrates cell distribution in our nested grid.
It denotes the cross section in the two dimension for simplicity. 

The gravitational potential is evaluated at the cell center
and is labeled as
\begin{equation}
\phi _{i,j,k} ^{(\ell)} \; = \;
\phi \, \lbrack x _{i} ^{(\ell)}, \, y _{j} ^{(\ell)},
\, z _{k} ^{(\ell)} \rbrack \; .
\end{equation}
Also the cell averaged density is labeled as
\begin{equation}
\rho _{i,j,k} ^{(\ell)} \; = \;
\rho \, \lbrack x _{i} ^{(\ell)}, \, y _{j} ^{(\ell)},
\, z _{k} ^{(\ell)} \rbrack \; .
\end{equation}

The Poisson equation,
\begin{equation}
4 \pi G \rho \; = \;
 \mbox{\boldmath$\nabla$}\cdot\mbox{\boldmath$g$} \;  ,
\end{equation}
and
\begin{equation}
\mbox{\boldmath$g$} \; = \; \mbox{\boldmath$\nabla$} \phi \; ,
\end{equation}
relates the gravitational potential and density,
where $ \mbox{\boldmath$g$} $ denotes the gravity.
When a cell is surrounded by cells of the same size,
we obtain the discrete Poisson equation of the second order
accuracy,
\begin{eqnarray}
4 \pi G \rho _{i,j,k} ^{(\ell)} & = &
\frac{g _{x,i+1/2,j,k} ^{(\ell)} \, - \, 
g _{x,i-1/2,j,k} ^{(\ell)}}{2^{-\ell+1} h} \; + \;
\frac{g _{y,i,j+1/2,k} ^{(\ell)} \, - \, g _{y,i,j-1/2,k} ^{(\ell)}}
{2 ^{-\ell+1} h} \nonumber \\ & \; & \; + \;
\frac{g _{z,i,j,k+1/2} ^{(\ell)} \, - \, g _{z,i,j,k-1/2} ^{(\ell)}}
{2 ^{-\ell+1} h} \; , \label{dpoisson1}
\end{eqnarray}
and
\begin{eqnarray}
g _{x,i+1/2,j,k} ^{(\ell)} & = & 
\frac{\phi _{i+1,j,k} ^{(\ell)} \, - \,
\phi _{i,j,k} ^{(\ell)}}{2 ^{-\ell+1} h} \label{gx} \; , 
\label{dpoisson2} \\
g _{y,i,j+1/2,k} ^{(\ell)} & = & 
\frac{\phi _{i,j+1,k} ^{(\ell)} \, - \,
\phi _{i,j,k} ^{(\ell)}}{2 ^{-\ell+1} h} \label{gy} \; , 
\label{dpoisson3} \\
g _{z,i,j,k+1/2} ^{(\ell)} & = & 
\frac{\phi _{i,j,k+1} ^{(\ell)} \, - \,
\phi _{i,j,k} ^{(\ell)}}{2 ^{-\ell+1} h} \label{gz} \; ,
\label{dpoisson4}
\end{eqnarray}
by the central difference as in case of the uniform grid.
Note that the gravity is evaluated on the cell surface.

Equations (\ref{dpoisson2}) through (\ref{dpoisson4}) need to be
modified near the grid boundary. 
We set the condition that the gravity evaluated in the coarse grid
is equal to the average gravity on the smaller cell surfaces, e.g.,
\begin{eqnarray}
g _{x,i+1/2,j,k} ^{(\ell)} & = & \frac{1}{4} \, 
\lbrack g _{x,2i+3/2,2j,2k} ^{(\ell+1)} 
\, + \, g _{x,2i+3/2,2j,2k+1} ^{(\ell+1)} \nonumber \\
& \; & 
\, + \, g _{x,2i+3/2,2j+1,2k} ^{(\ell+1)} \, 
\, + \, g _{x,2i+3/2,2j+1,2k+1} ^{(\ell+1)} \rbrack \; .
\label{gauss}
\end{eqnarray}
This ensures that our solution satisfies the Gauss's 
theorem.  When we sum up the normal component of the gravity 
over the surface of a given volume, it equals to the
mass contained in the volume multiplied by $ 4 \pi G $.
\begin{equation}
\sum _{\rm surface} \mbox{\boldmath$g$} \cdot d\mbox{\boldmath$S$}
\; = \; 4 \pi G \, \sum _{\rm volume} \rho \, dV \; .
\end{equation}
In other words, the \lq\lq gravitational field line'' 
never ends at the grid level boundary.  This is also equivalent
to set the Neumann condition at the grid level boundary.

When the cell surface is on the grid level boundary,
we interpolate $ \phi $ in the coarse grid to evaluate
the gravity across the cell surface.
As an example we show the interpolation formula to compute
the gravity in the $ x $-direction in the following.
When $ i \, = \, N/2 \, - \, 1$, we evaluate 
\begin{equation}
g _{x,i+1/2,j,k} ^{(\ell)} \, = \, \frac{\phi _{i+3/2,j,k} ^{*(\ell)}
\, - \, \phi _{i,j,k} ^{(\ell)}}{(3/2) \times 2 ^{-\ell+1} h}  \;,
\label{interpolation1}
\end{equation}
where $ \phi _{i+3/2,j,k} ^{*(\ell)} $ denotes the gravitational
potential at $ (x, \, y, \, z) $ = 
$ \lbrack x _{i+3/2} ^{(\ell)} \, \equiv \,
x _{(i+1)/2} ^{(\ell-1)}, \, y _j ^{(\ell)}, \, z _k ^{(\ell)} 
\rbrack $.  It is obtained by the interpolation on the diagonal
in the coarse cell surface.
When $ j $ and $ k $ are odd numbers, it is evaluated to be
\begin{equation}
\phi _{i+3/2,j,k} ^{*(\ell)} \; = \; 
\frac{3}{4} \, \phi _{(i+1)/2,(j-1)/2,(k-1)/2} ^{(\ell-1)} \;
+ \;
\frac{1}{4} \, \phi _{(i+1)/2,(j+1)/2,(k+1)/2} ^{(\ell-1)} \; . 
\label{interpolation2}
\end{equation}
Equation (\ref{interpolation2}) uses the gravitational potential
at the both ends of the diagonal for the linear interpolation.
This interpolation is not unique; we can use the bi-linear
interpolation in the coarse cell surface.  The result depends
little on the interpolation adopted. 
The cell number at the both ends of the diagonal
depends on the even-odd parity of $ j $ and $ k $.
Equation (\ref{interpolation2}) should be modified 
appropriately when either $ j $ or $ k $ are even.

Equations (\ref{gx})-(\ref{gz}) are central difference,
our difference equations are the second order accurate
except across the grid level boundaries.
Since equations (\ref{interpolation1}) and (\ref{interpolation2})
are the first order accurate, our difference equations
are only the first order accurate across the boundaries.

We need the gravity at the cell center in the hydrodynamical
computation.  The gravity at the cell center is evaluated
to be the average of those at the opposing cell surfaces,
e.g.,
\begin{equation}
g _{x,i,j,k} ^{(\ell)} \; = \; 
\frac{1}{2} \, \left\lbrack 
g ^{(\ell)} _{x,i+1/2,j,k} \, + \,
g ^{(\ell)} _{x,i-1/2,j,k} \right\rbrack \; . \label{c-center-g}
\end{equation}
We examine the accuracy of the gravity at
the cell center in \S 4.

\section{Iterative Method for Solving Difference Equation}

Our difference equation can be solved with a simple 
point Jacobi iteration or red-black Gauss-Seidel iteration
but with too many times iteration of the order of $ N ^2 $.
Thus we adopt the multi-grid iteration for to accelerate
the convergence.  We employ the Full Multi-Grid
(FMG) scheme, i.e., the algorithm shown in \citet{press91}
in our paper.  Our numerical procedures are rather complicated
partly because the FMG scheme is complicated and partly 
because our nested grid is complex.  Thus, we first
outline our scheme in the following.  Detailed numerical
procedures are shown later.  

Essence of the multi-grid iteration is to obtain a better
initial guess for finer grids from an approximate solutions 
on coarser grids \citep[see, e.g.,][for the basic of 
multi-grid iteration]{wesseling92,briggs00}.  
When grids are coarse, the computation
cost is less.  When interpolated for a fine grid, the
solution on a coarse grid is a good initial guess and
a better solution on the fine grid is obtained by only
a few times iterations.

As a coarse grid, we use another nested grid which covers
the same computation volume with a fewer $ N $ and
larger $ h $.  More specifically we use the nested grids
of $ (N, \, h) $ = $ ( 2 ^{n-1}, \, 2 h_0) $, 
$ ( 2 ^{n-2}, \, 2 ^2 h _0) $, ...., and 
$ (2 ^2, \, 2 ^{n-2} h _0 ) $,
as temporarily working nested grids for computation on
the nested grid of $ (N, \, h ) $ = $ (2^n, \, h _0) $
\citep[see Chapter 9 of][for the reasoning on our choice
of coarsening]{briggs00}.
Furthermore, we introduce two uniform grids one of which
has $ 2^3 $ cells and the other of which has only 1 cell.
Both the uniform grids covers the whole computation domain
which is covered only with the largest grid in the 
nested grid.  
Introduction of these temporarily working grids 
increases the total number of cells in our computation only 
a factor of 8/7.  Accordingly the computation on these
working nested grids is only a minor fraction.

We start from an exact solution on the uniform grid
having only one cell.  The solution is used as an
initial value to obtain the solution on the other uniform
grid.  The obtained solution is used to obtain the
solution on the nested grid of $ (N, \, h) $ =
$ (2^2, \, 2 ^{n-2} h _0) $.  This solution is
obtained by the Successive Over Relaxation 
\citep[SOR, see, e.g.][for the basic of SOR]{press86}.
The computation cost is very small since this nested
grid contains only $ 64 \, \ell _{\rm max} $ cells.
We obtained a solution on the coarsest grid after
$ 4 \ell _{\rm max} $ times iteration in a typical model.
The iteration should reduces the residual 
by a factor of $ 10 ^{-3} $, since $ 4 p N _1 / 3 $ times
iteration reduces the residual by a factor of $ 10 ^{-p} $
in SOR \citep{press86} when the grid has $ N _1 $ cells in
one dimension.  By interpolating the
solution we obtain the initial data for iteration on
the grid of $ (N, \, h) $ = $ (2 ^3, \, 2 ^{n-3} h _0) $.
After a few times (typically two times) red-black
Gauss-Seidel iteration, we obtain an approximate solution
on the fine grid.

\begin{figure}
\epsscale{0.8} 
\plotone{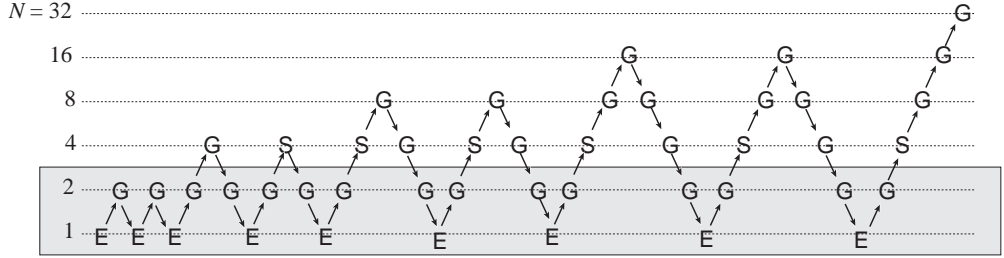}
\caption{Schematic diagram of FMG scheme.  The symbols G, S, and E
denote the red-black Gauss-Seidel iteration, SOR, and solving an exact
solution, respectively. The arrows directing right upward and right
downward denote interpolation and restriction, respectively.  In the
hatched levels, the operations are done only in the $\ell = \ell_{\rm
max}$.  }
\label{mg.eps}
\end{figure}

We obtain an approximate solution on a finer nested grid
successively by combination of interpolation, the red-black
Gauss-Seidel iteration and restriction (averaging).
We use the bilinear formula to interpolate the solution
on a coarse grid for that on a fine grid. 
On the other hand we use the full weighting 
(the simple volume average) to restrict the solution on a
fine grid to that on a coarse grid.
We perform these operations on the order of the FMG.
The schedule of the operations are schematically shown
in Figure \ref{mg.eps}.  The arrows directing right upward denote
interpolation while those directing right downward do
restriction.  
The symbol, G, denotes the red-black Gauss-Seidel iteration while
the symbol, S, dose SOR.
At the points of the symbol, E, an exact solution is given.
The red-black Gauss-Seidel iteration is performed twice
both after interpolation and interpolation in a typical
model.  In other words, the number of pre- and post-iterations
is two in our computation.

Each restriction is performed by simply taking the average
in the cells involved in a coarse cell.
Restriction is done from fine 
($ \ell \, = \, \ell _{\rm max} $) to coarse 
($ \ell \, = \, 1 $). 

Each interpolation is performed in the following procedures.
\begin{enumerate}
\item Fill the data in the region covered with a finer
grid by the average value in the finer grid.  This
averaging is performed successively from $ \ell \, = \,
\ell _{\rm max} \, - \, 1 $ to $ \ell \, = \, 1 $.
\item Perform linear interpolation of three dimensional
data successively from $ \ell \, = \, 1 $ to 
$ \ell \, = \, \ell _{\rm max} $.  The data on the
boundary is given from the coarser grid.  
\end{enumerate}

Each red-black iteration is performed in the following
procedures.
\begin{enumerate}
\item Obtain the gravitational potential just outside
the grid level boundary using Equation (\ref{interpolation2}).
\item Evaluate the gravity, \mbox{\boldmath$g$}, on
the cell boundary using Equations (\ref{gx}), (\ref{gy}),
(\ref{gz}) and (\ref{interpolation1}).
\item Replace the gravity at the grid level boundary with
the fine grid using Equation (\ref{gauss}).
\item Compute the residual in the red 
($ i \, + \, j \, + \, k $ = odd) cells and 
update the data there.
\item Repeat procedures 1 -- 3.
\item Compute the residual in the black 
($ i \, + \, j \, + \, k $ = even) cells and 
update the data there.
\end{enumerate}
The above procedures are counted as a cycle of the 
red-black Gauss-Seidel iteration.

The boundary condition was set only on the boundaries of
the largest grid.  We can use either the Dirichlet boundary 
(setting $ \phi $ on the boundary) or the Neumann boundary
(setting $ \mbox{\boldmath$g$} $).  In the following
we use the Neumann boundary.  The gravity on the boundary
was computed by the multi-pole expansion.

\section{Sample Computation}

We evaluate the efficiency of the above described
algorithm with emphasis on the accuracy, convergence and
computation time (load).  In our numerical computation
we take the unit, $ G \, = \, 1 $, for convenience.
All the numerical computations were done in the double
precision.

\subsection{Accuracy}
\label{accuracy}
First we present an example in which two uniform 
density spheres are placed near the center of the
nested grid.   The central positions,  
radii, and masses of the spheres are listed in Table 1.
We obtained the gravitational potential for this mass
distribution with the method mentioned in the previous
sections.  We used the nested grid of
$ N \, = \, 64 $, $ L \, = \, 1 $, and $ \ell _{\rm max} \, = \, 5$.
This example provides us 
the accuracy of our scheme.

\begin{figure}
\epsscale{0.3}
\plotone{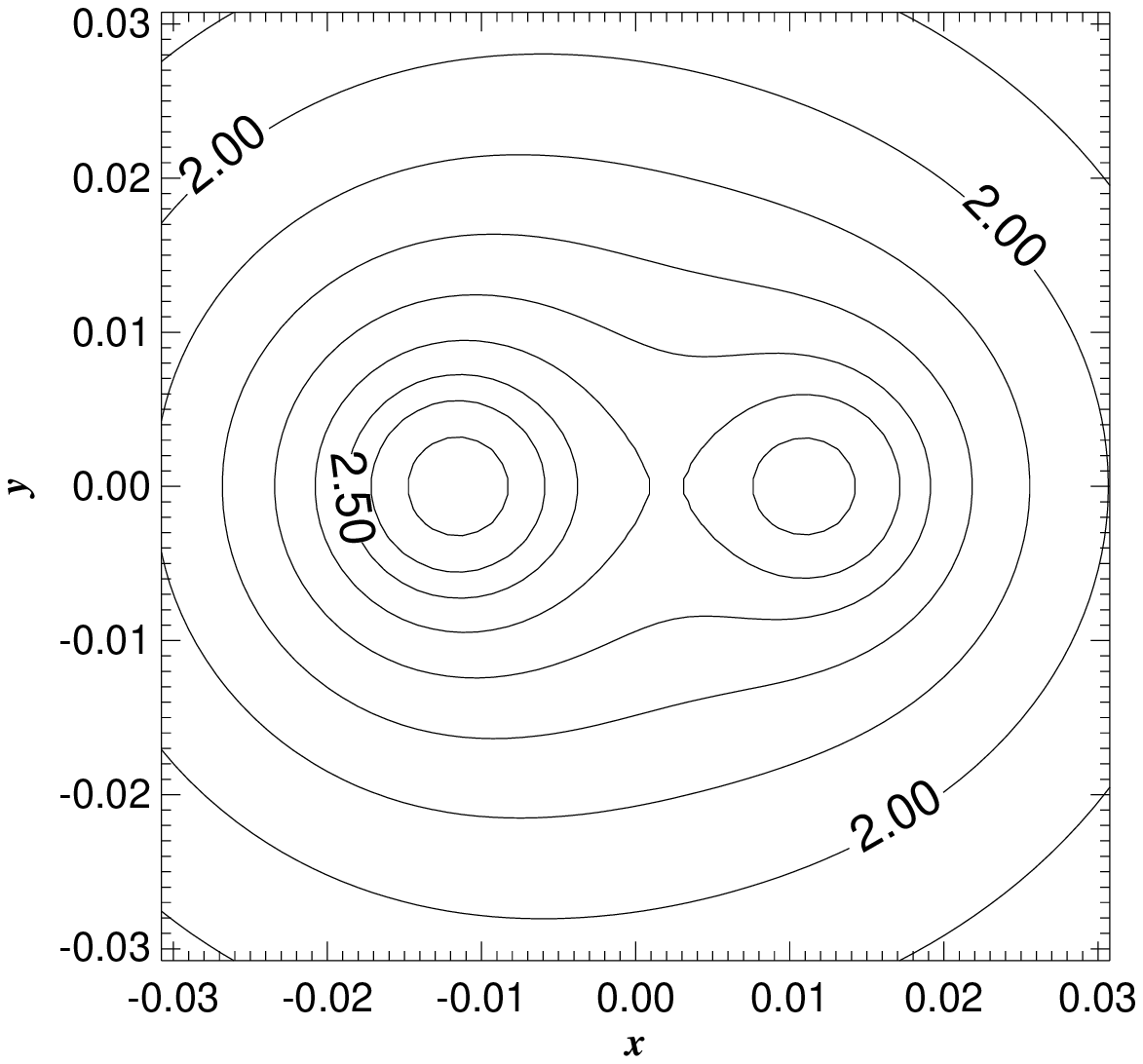}
\plotone{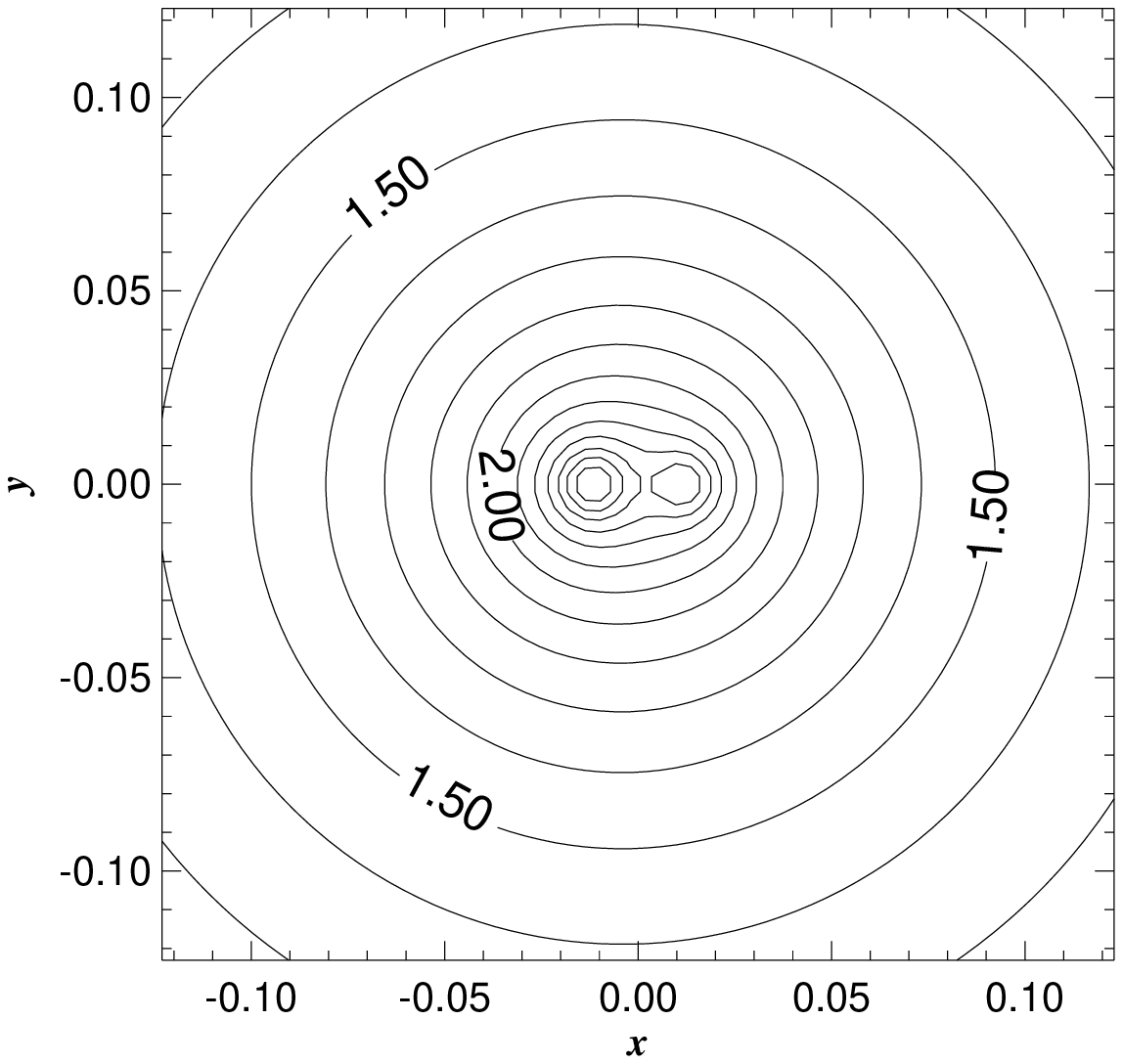}
\plotone{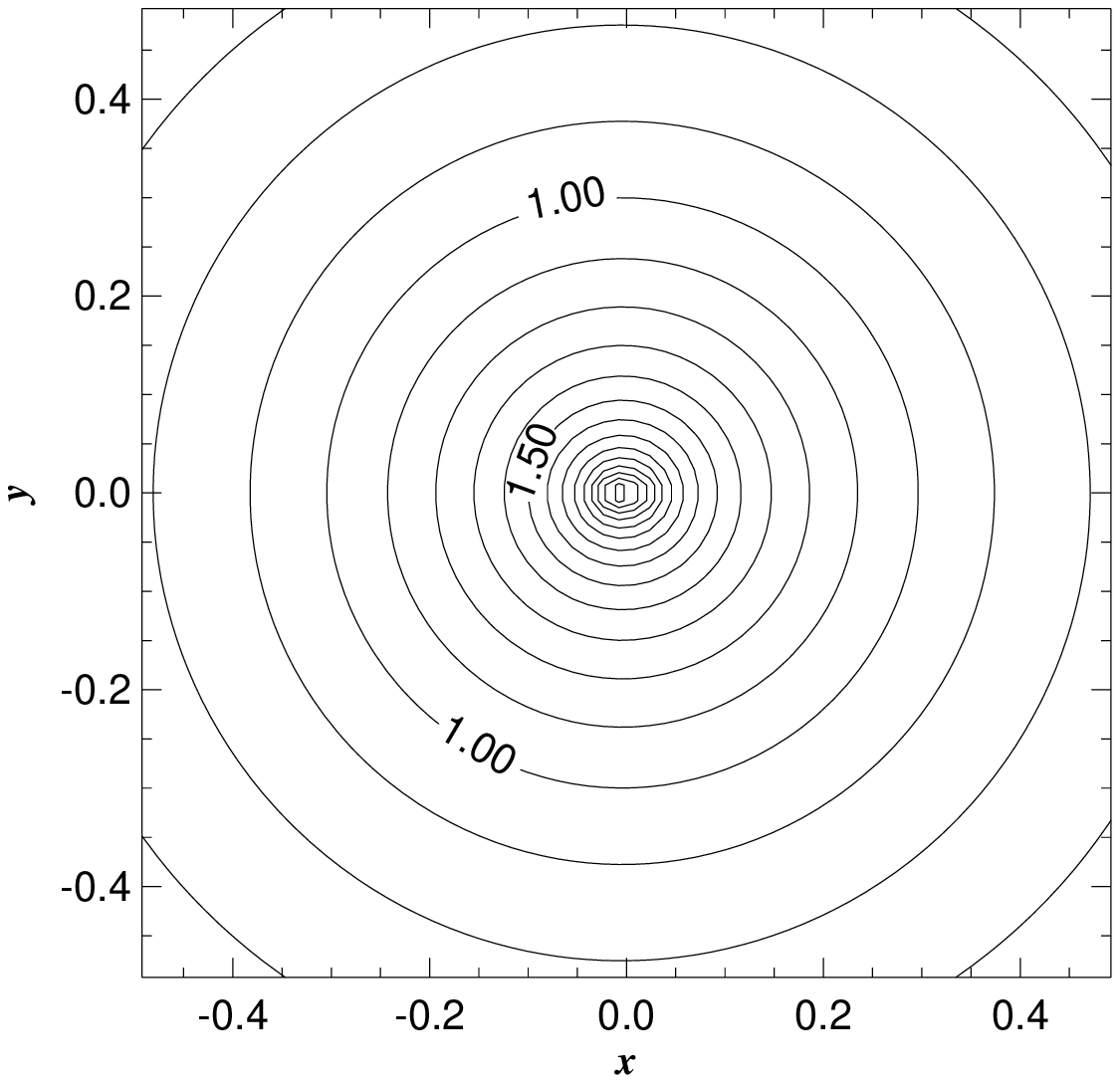}
\caption{Example of our numerical solution for gravitational
potential of binary stars.  Each component star of the binary is
assumed to be a sphere of the uniform density.  
The centers, radii, and densities are summarized in Table 1.
The contours denote the computed gravitational
potential.  It is measured in the plane of $ z $ = 1/2048 at
the level of $ \ell \, = \, 5 $ in panel ({\it a}),
in the plane of $ z $ = 1/512 at the level of 
$ \ell \, = \, 3 $ in panel ({\it b}), and
in the plane of $ z $ = 1/128 at the level of 
$ \ell \, = \, 1 $ in panel ({\it c}).
The contours are set in the equal interval in the logarithmic
scale, $ \Delta \log \vert \phi \vert \, = \, 0.1 $.
The labels denote the values of $ \log \, \vert \phi \vert $.}
\label{binary4.eps}
\end{figure}

We started the computation from the initial 
guess, $ \phi \, = \, 0 $.  The residual decreased by more
than a factor of several hundreds by each FMG
iteration.  After 5 times iteration, the residual was of
the order of the round off error 
($ \approx \, 10 ^{-15} $ in the potential) 
as will be shown in \S\S \ref{convergence}.
The obtained gravitational potential is shown in 
Figures \ref{binary4.eps}{\it a}, {\it b}, and {\it c}.  Each panel denotes the
contours of the gravitational potential measured in the
midplane.  As shown in Figure \ref{binary4.eps}{\it a}, the gravitational potential
$ \phi $ has two local minima of which centers are close to
the those of two spheres.  It is well approximated by  
the $1/r$ potential in the region far from these spheres.
These features ensure that the obtained solution is 
qualitatively correct.

\begin{figure}
\epsscale{0.3}
\plotone{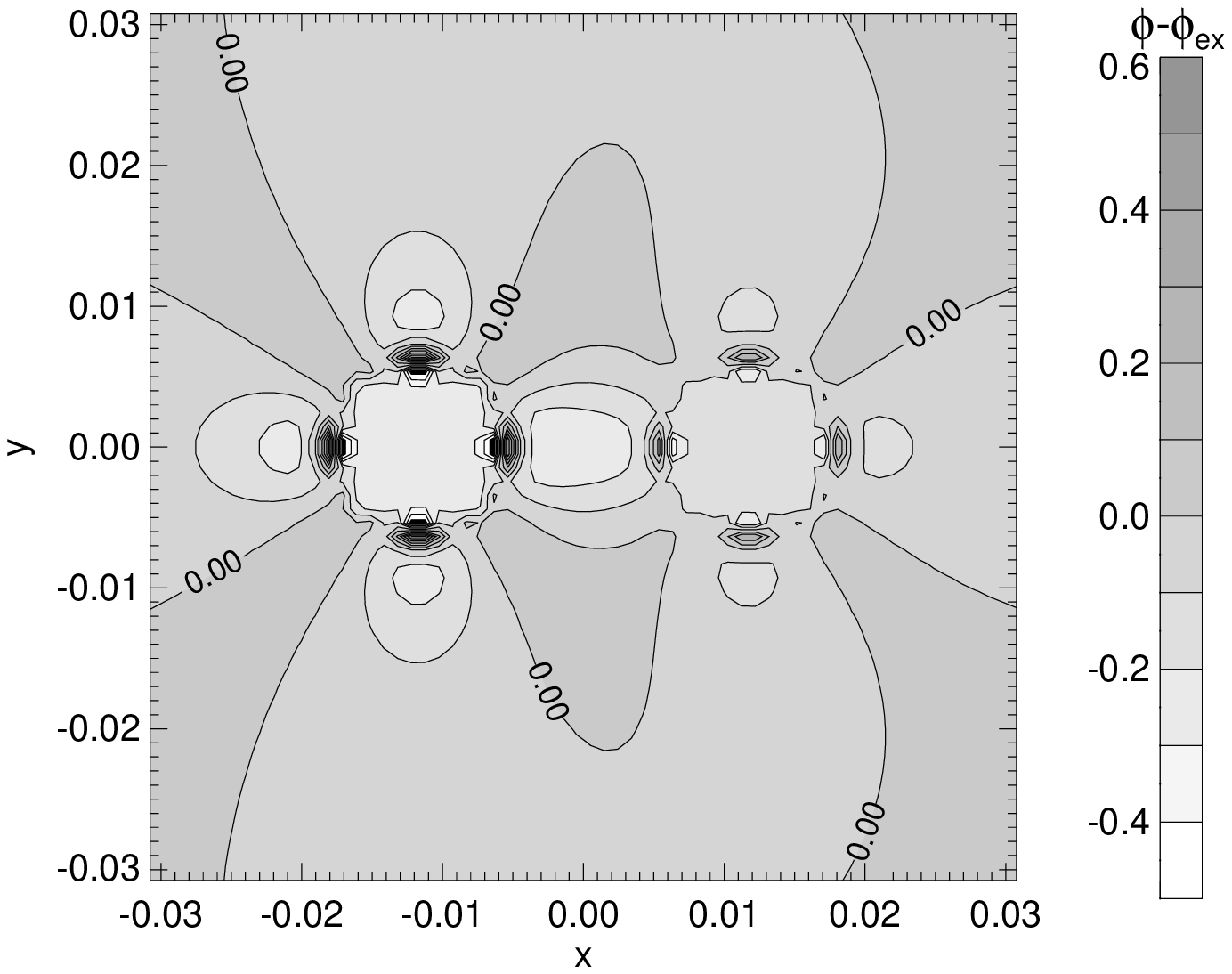}
\plotone{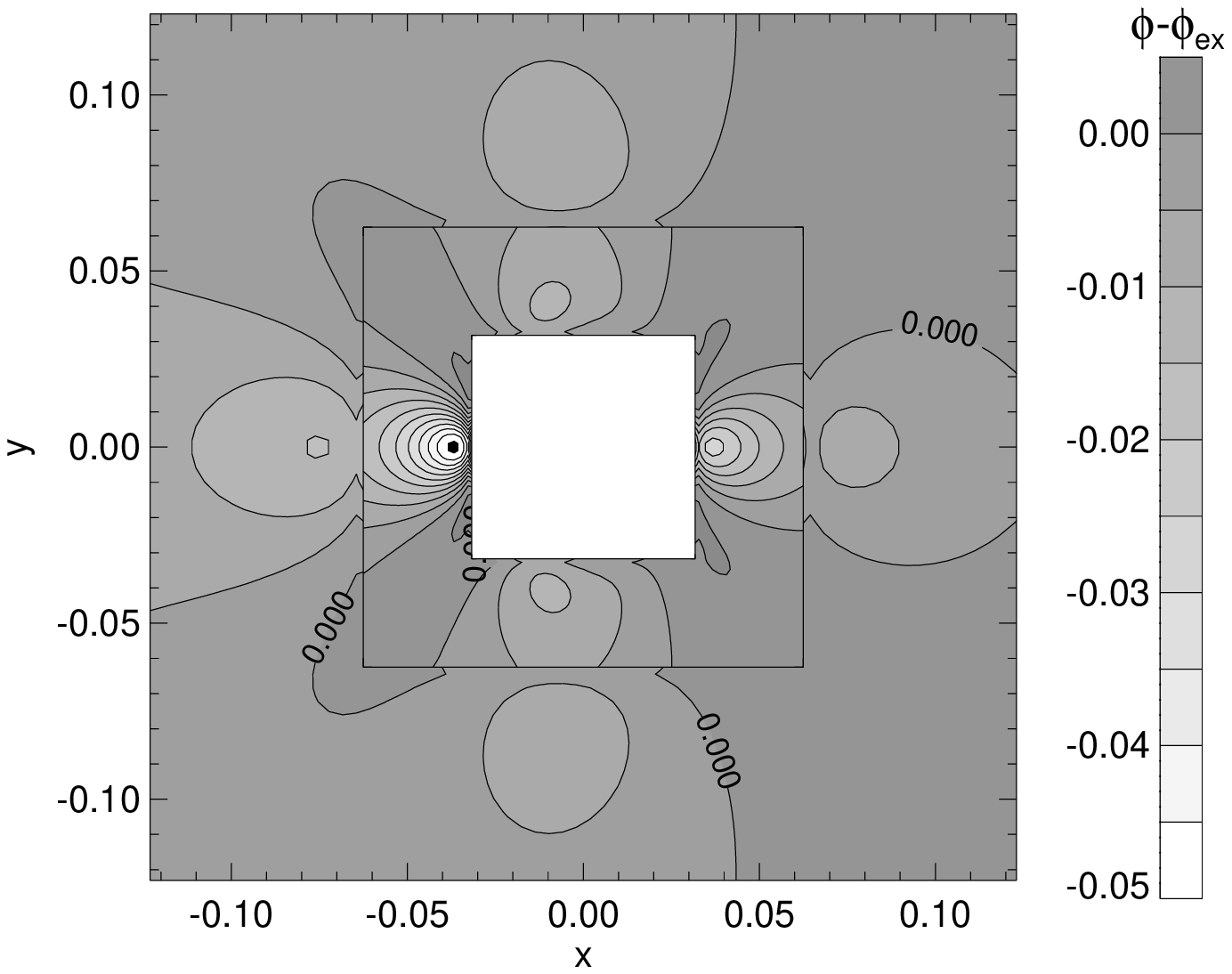}
\plotone{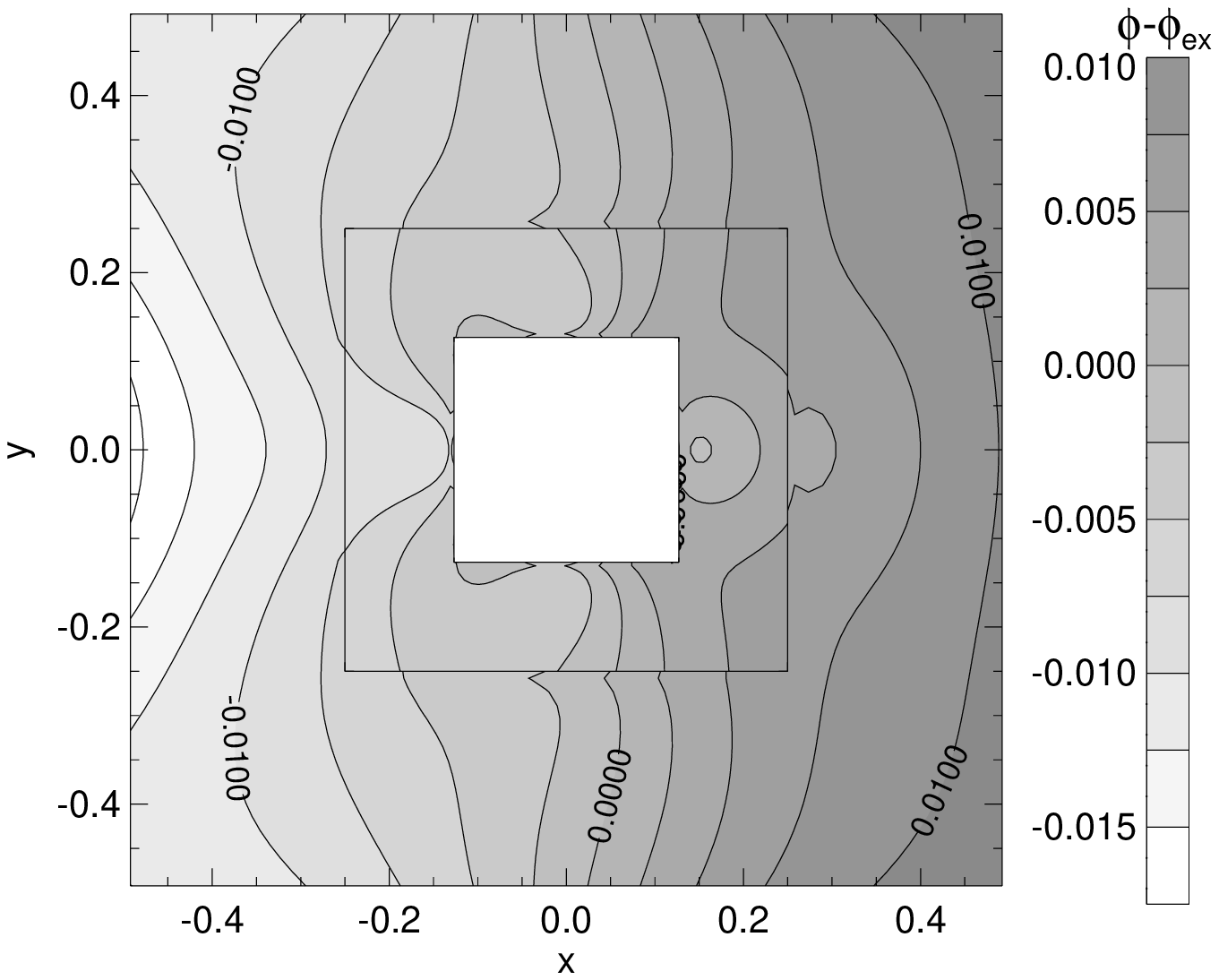}
\caption{Errors of gravitational potential due to difference scheme
for the example shown in Figure \ref{binary4.eps}.
The grayness and contours denote the difference form  analytic
solution ($\Delta \phi 
= \phi - \phi_{\rm ex}$).
Panel ({\it a}) denotes the difference on $ z \, = \, 1/2048 $ in the grid
level of $ \ell \, = \, 5 $.  Panel ({\it b}) denotes that on $ z \, = \,
1/1024 $ and 1/512 in the grid levels of $ \ell \, = \, 4 $ and 3,
respectively.  Panel ({\it c}) denotes that on $ z \, = \, 1/256 $ and 1/128
in the grid levels of $ \ell \, = \, 2 $ and 1, respectively.  }
\label{error4.eps}
\end{figure}

We compare the above solution with the exact analytic
solution to examine the accuracy of our difference
scheme.   Figure \ref{error4.eps}{\it a}, {\it b}, and {\it c} denote the difference between the
numerical and analytic exact solutions for the example shown in
Figure \ref{binary4.eps}.  We do not show the
difference in the region covered with a finer grid, since
the numerical solution has no practical use there.
Our numerically obtained $ \phi $ has an arbitrary offset,
since we used the Neumann boundary.  The offset was chosen
to minimize the difference between the numerical and
analytic solution at $ z \, = \, 1023/2048 $.
On the grid of $ \ell \, = \, 5 $, the difference is large 
($ \vert \Delta \phi \vert \, = \, 0.6 $) in the cells located 
near the surfaces of the spheres as shown in Figure
\ref{error4.eps}{\it a}.  
This error is inevitable 
since it is due to the finiteness of our grids.  It is, however,
much smaller than the absolute value of the gravitational potential,
of which maximum is $ \max \, \vert \phi \vert \, = \, 552.0 $.
The difference from the exact solution is typically as large
as $ \vert \Delta \phi \vert \, = \, 0.1 $ in the region covered
with the grid of $ \ell \, = \, 5 $.  

The difference from the analytic solution
is less than $ \vert \Delta \phi \vert \, < \, 0.016 $ in the
region covered with the grid of $ \ell \, = \, 3 $ as shown
in Figure \ref{error4.eps}{\it b}.  The difference is relatively large near the
grid level boundary between $ \ell \, = \, 3 $ and 4.  This is 
most likely due to the evaluation of the gravity at the
grid level boundary.  Our numerical scheme is the second
order accurate except at the grid level boundaries since
we employ the center difference.  The gravity at the
grid level boundary is only the first order accurate [see
Eqs. (\ref{interpolation1}) and (\ref{interpolation2})].

On the grid of $ \ell \, = \, 1 $, the
error is global; it is negative in the left ($ x \, < \, 0 $)
and positive in the right ($ x \, > \, 0 $).  This error
comes from the boundary condition.  We applied the Neumann 
condition for the outer boundary.  The gravity at the
boundary is evaluated by the multi-pole expansion in which
we take account of the dipole, quadruple, and 
octuple moments as well as the total mass.  These higher order
moments and total mass are evaluated from the
density distribution on the coarse grid of $ \ell \, = \, 1 $.
Thus even the dipole moment is seriously underestimated in
this example. This underestimation is dominant in the error
on the coarse grid $ \ell \, = \, 1 $.
The quadruple and octuple moments are practically
not taken into account on the boundary in the example.  
If we apply a better outer boundary condition, the error will
be reduced.

\begin{figure}
\epsscale{0.3}
\plotone{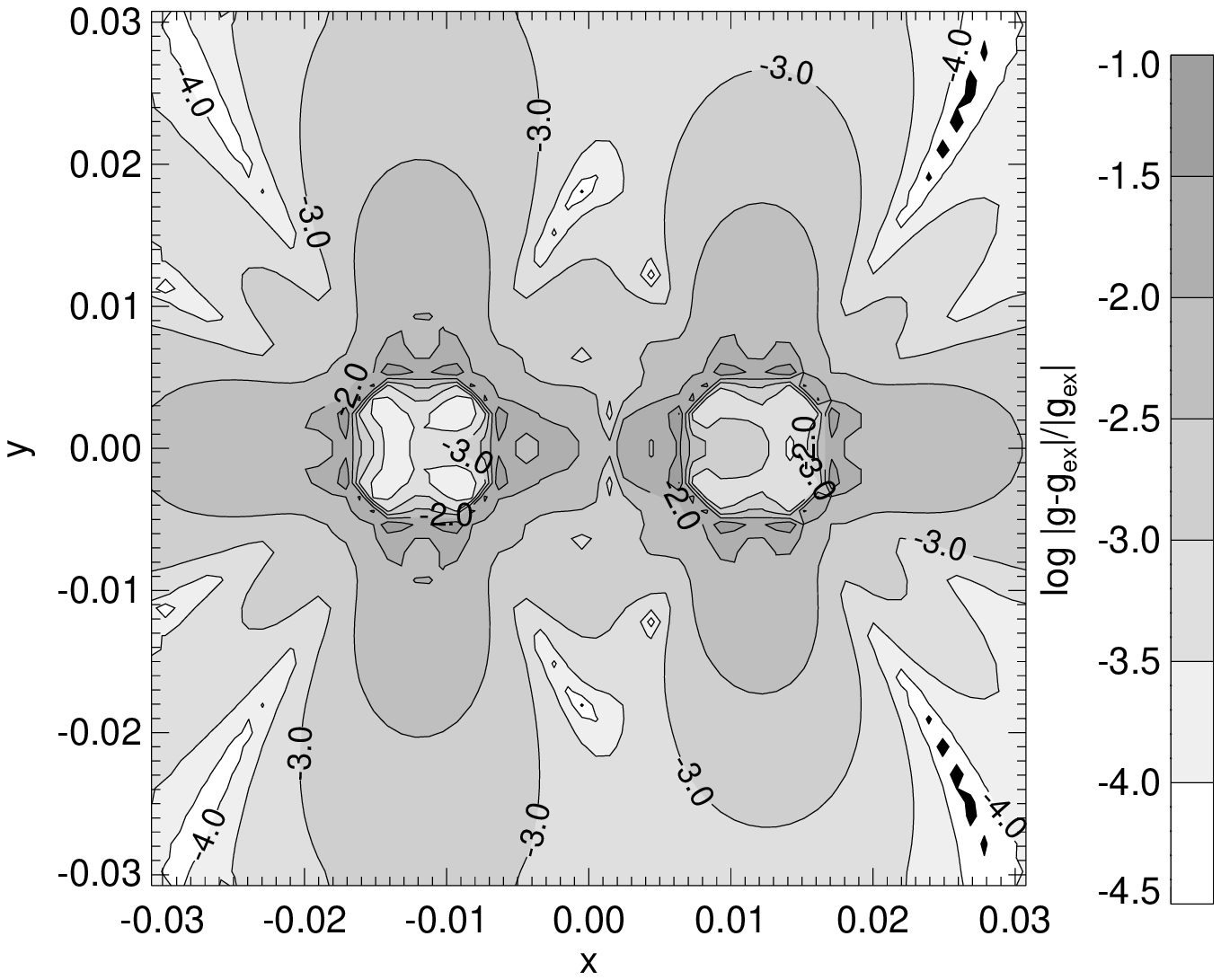}
\plotone{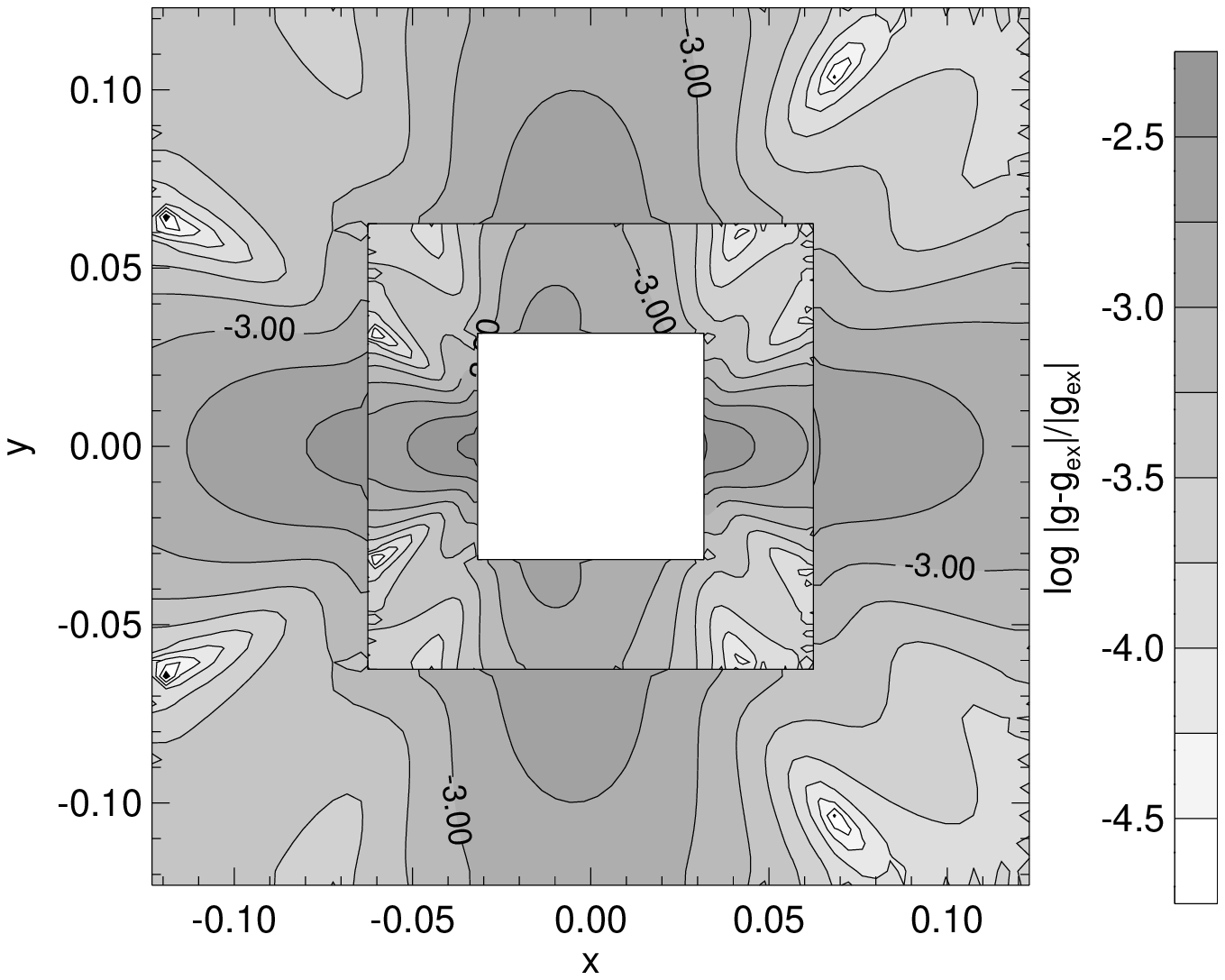}
\plotone{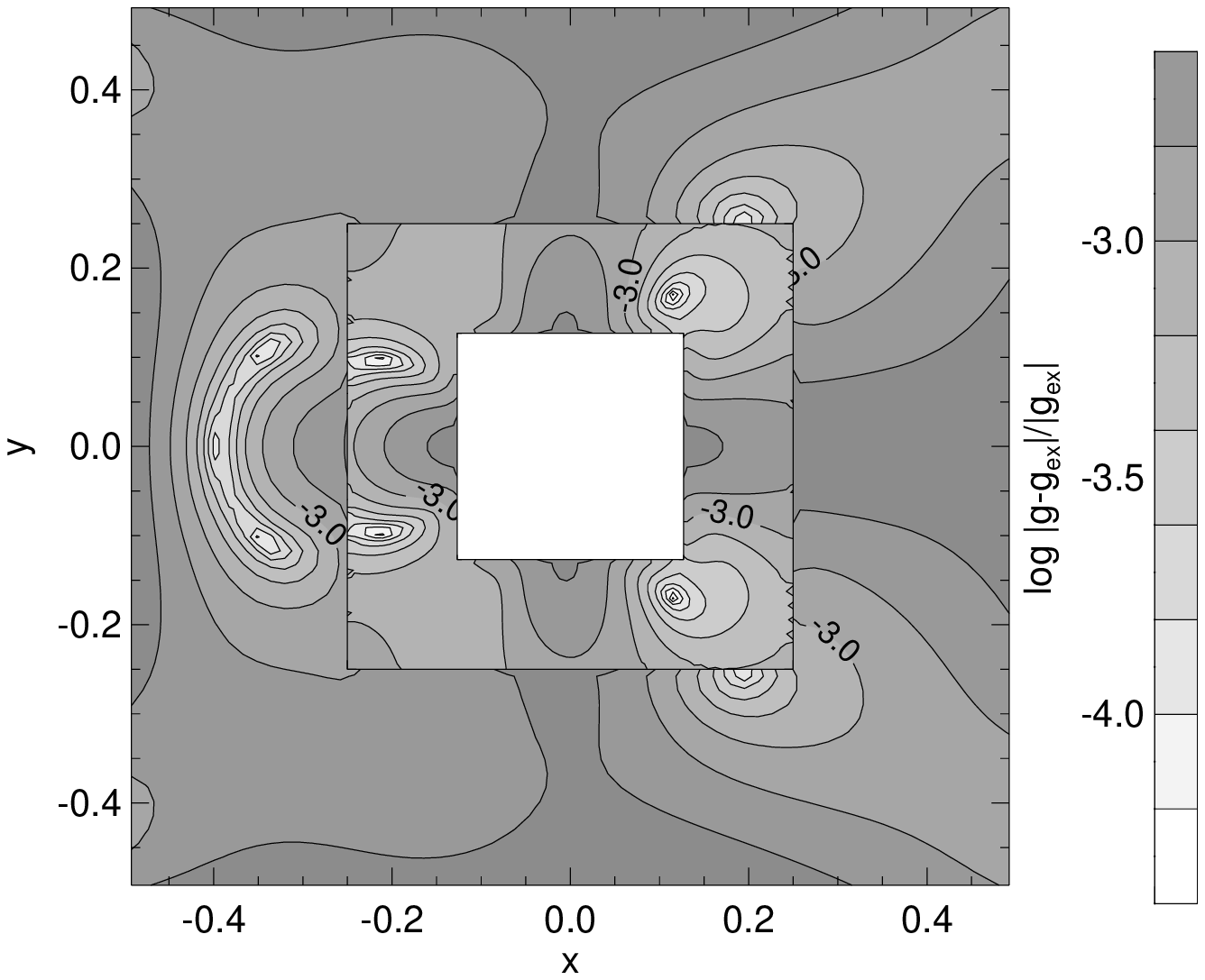}
\caption{
Same as Figure \ref{error4.eps} but for 
relative errors of gravitational force
($| \mbox{\boldmath $g$} - \mbox{\boldmath $g$} _{\rm ex} | /
|\mbox{\boldmath $g$} _{\rm ex} | $) in logarithmic scale.
}
\label{error-g.eps}
\end{figure}

Next we examine the accuracy of the gravity at the cell
center, $ \mbox{\boldmath$g$} _{i,j,k} $, defined by
Equation (\ref{c-center-g}).
Figure \ref{error-g.eps} is the same as Figure \ref{error4.eps} but for 
the relative error,
$ \vert \mbox{\boldmath$g$} \, - \, \mbox{\boldmath$g$} _{\rm ex} 
\vert / \vert \mbox{\boldmath$g$} _{\rm ex} $,
where $ \mbox{\boldmath$g$} _{\rm ex} $ denotes the exact
gravity obtained analytically.  The relative error is large
($ \sim 0.1 $) in the cells located on the surfaces of the 
spheres.  It is much less than $ 10 ^{-2} $ in most cells.
This small error ensures hat our scheme reproduces not only
the $ 1/r $ potential but also the quadruple moment of the
binary.  The error is moderately large ($ \sim 0.02 $) 
in the cells adjacent
to the grid level boundary.  It is fairly small except
in the cells located on the surfaces of the spheres.  
The large error on the surfaces of the spheres are due to
the fact that even the finest grid is too coarse to resolve
the surface sharply and not serious in practical use.
In most hydrodynamical simulations, a self-gravitating
gas has a more or less smooth density distribution
in the finest grid.

\begin{figure}
\epsscale{0.4}
\plotone{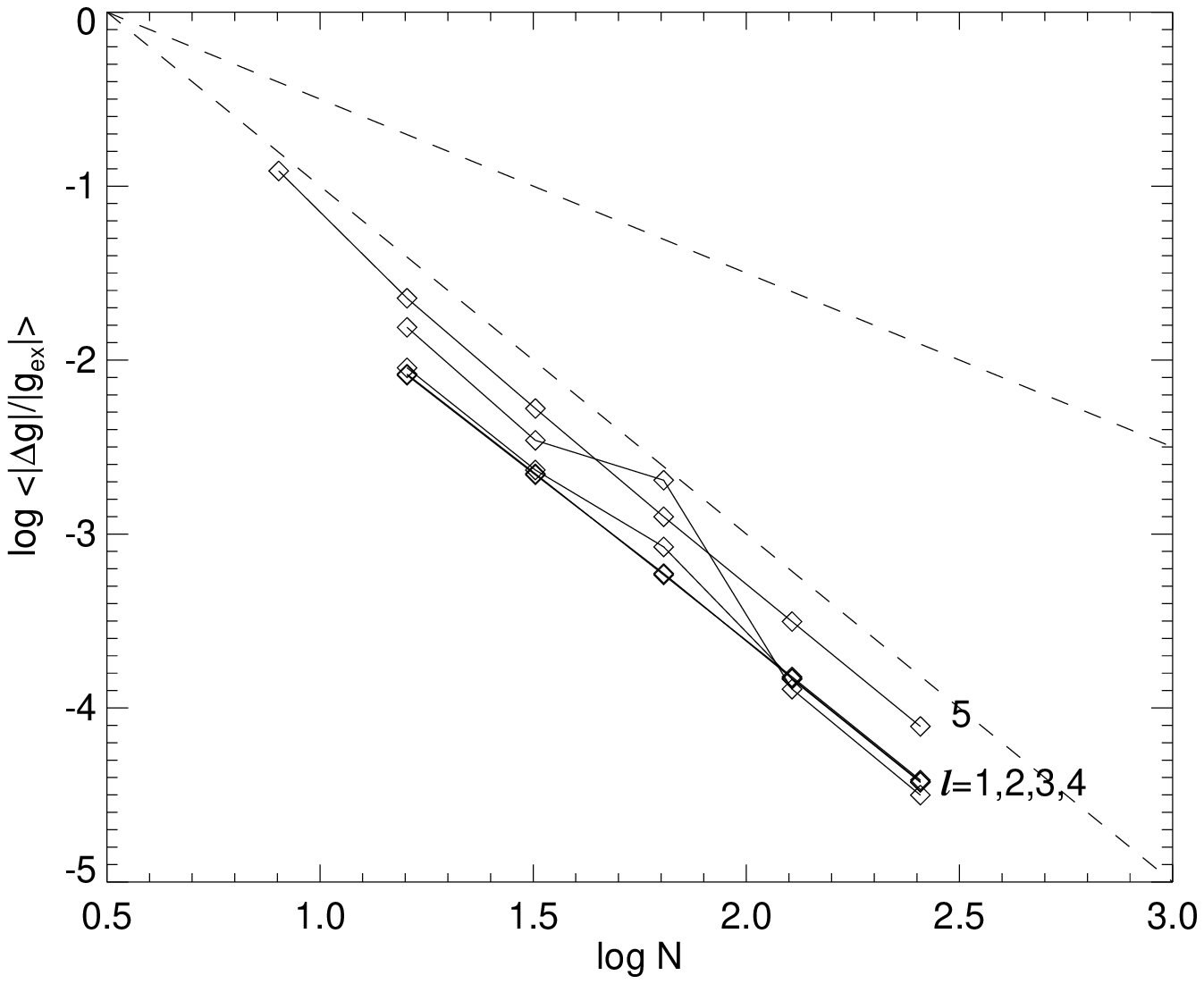}
\plotone{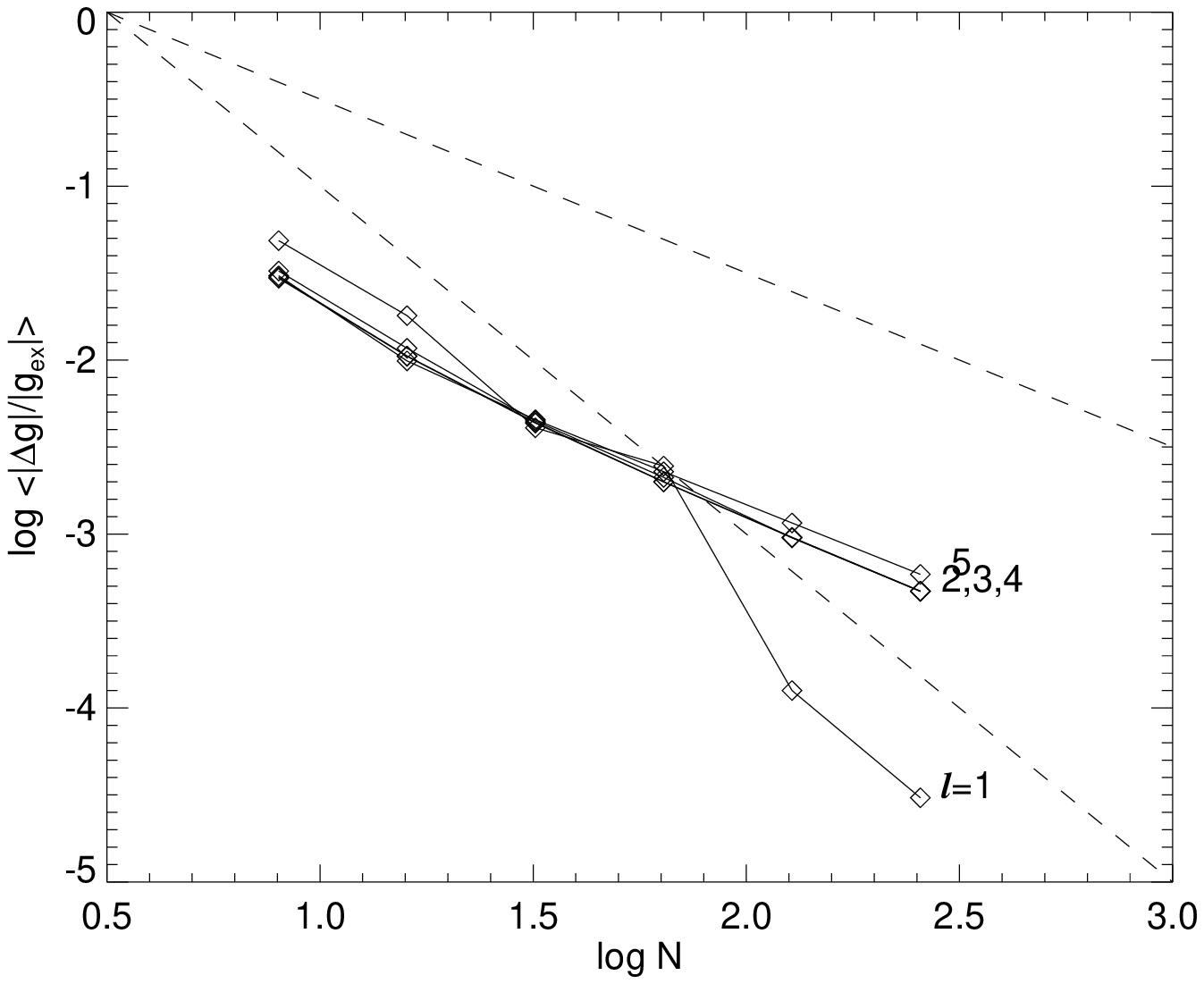}
\plotone{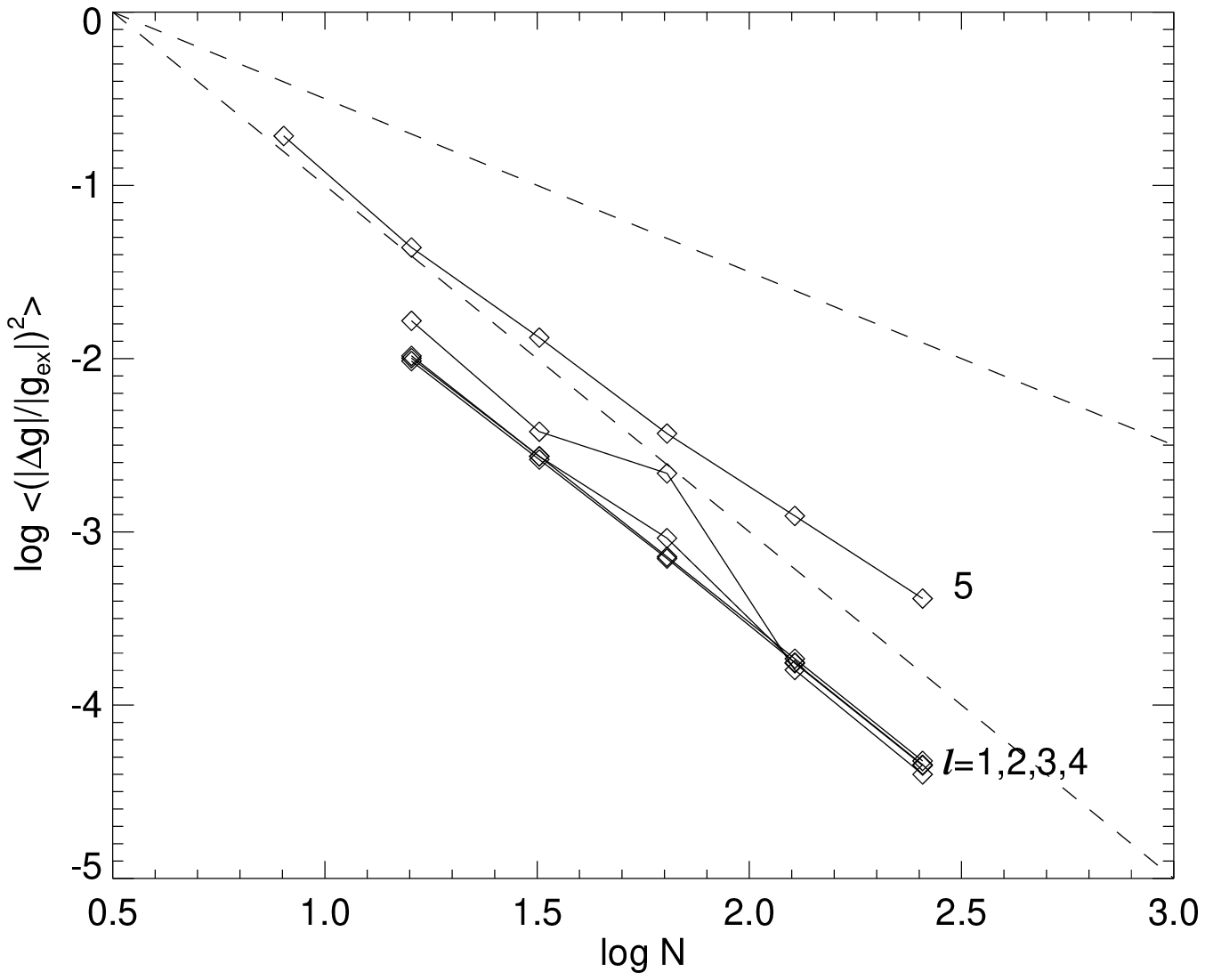}
\plotone{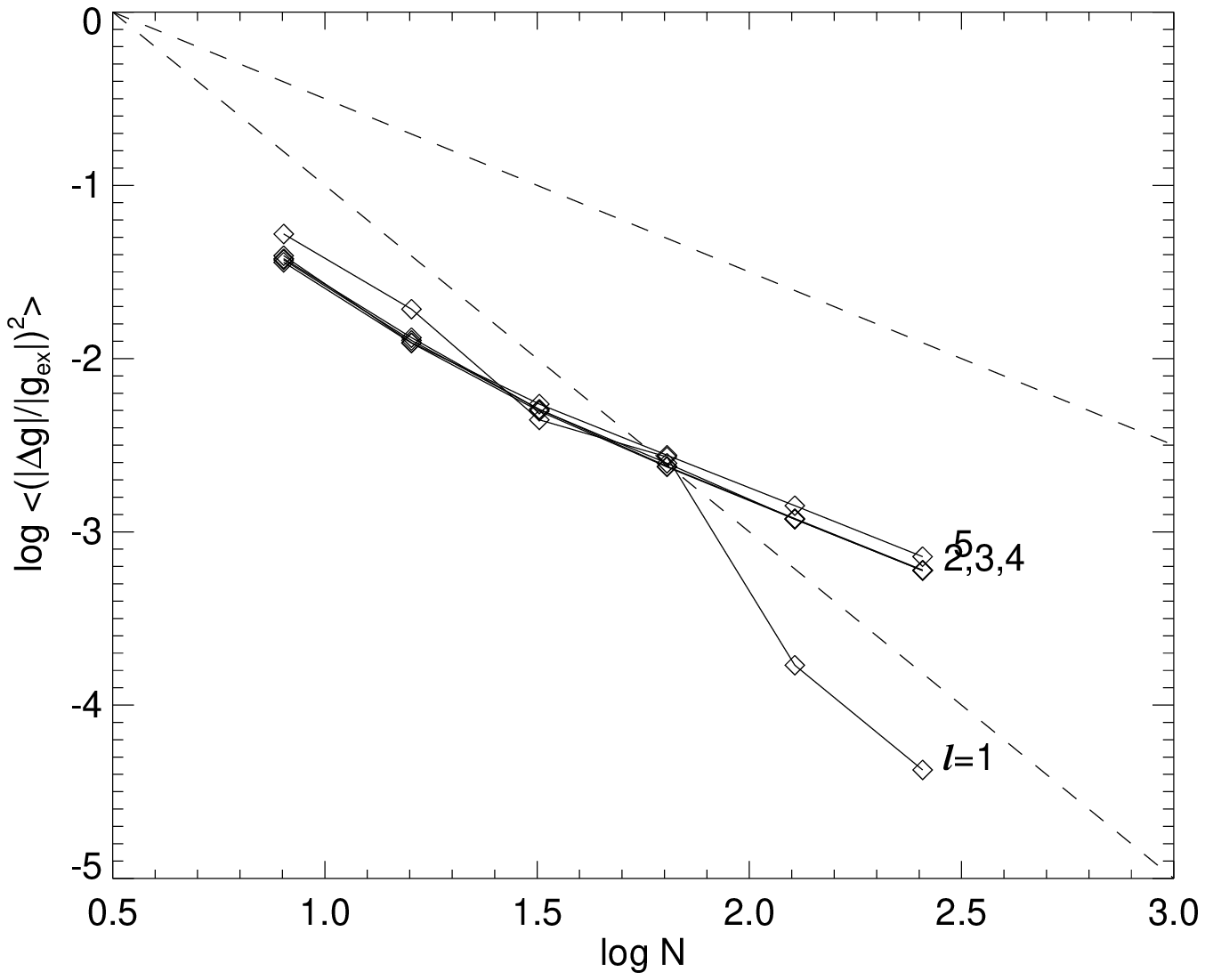}
\plotone{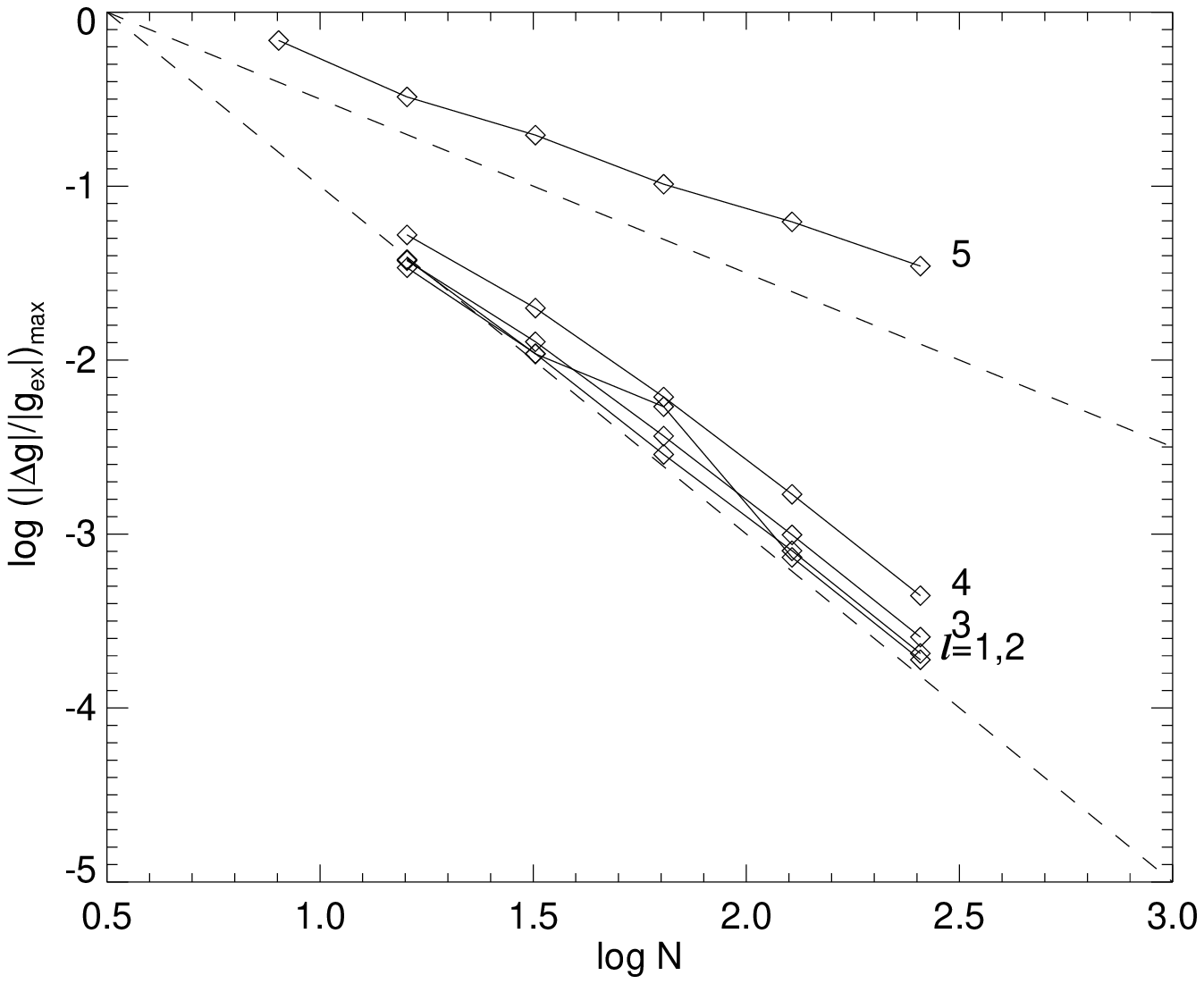}
\plotone{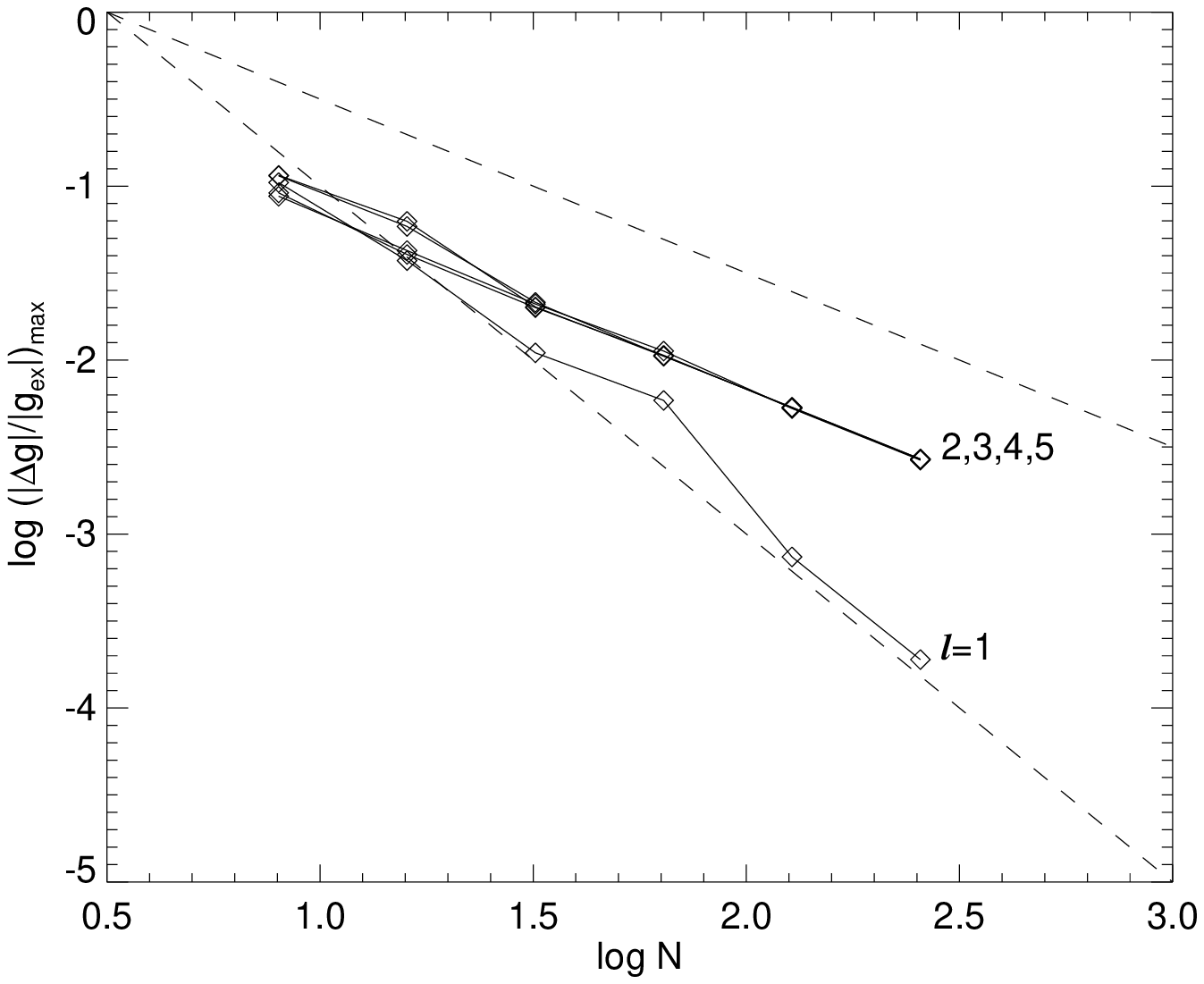}
\caption{
Errors of gravitational force as a function of $N$.
The errors are evaluated as ({\it a} and {\it b}) simple averages, 
({\it c} and {\it d}) root mean squares,
and ({\it e} and {\it f}) maximum
values of relative errors of gravitational force ($| \mbox{\boldmath
$g$} - \mbox{\boldmath $g$} _{\rm ex} | / 
|\mbox{\boldmath $g$} _{\rm ex} | $).
Panels ({\it b}), ({\it d}), and ({\it f}) denote the errors for cells on grid level
boundaries while panels ({\it a}), ({\it c}), and ({\it e}) 
denote the errors for the
other cells.
Each solid curve denotes 
the error for each grid level ($\ell = 1 - \ell_{\rm max}$) as
labeled in the panels.
The dashed lines denote $ 10 ^{0.5} \, N^{-1}$ and
$10 \, N^{-2}$ for comparison.
}
\label{norm.eps}
\end{figure}

To evaluate the relative error in the gravity quantitatively,
we measured the simple average, the root mean
square average, and the maximum value.  Figure \ref{norm.eps} 
show them
as a function of $ N $.  The average and maximum are 
computed separately for cells adjacent to the grid level 
boundary and for the rest cells.  Each curve denotes
the value at the grid of each level.  
Panels {\it b}, {\it d}, and {\it f} are for the cells 
adjacent to the grid level boundary while panels
{\it a}, {\it c}, and {\it e} are for the rest cells.
Panels {\it a} and {\it b} show the simple average.
Panels {\it c} and {\it d} show the root mean square average.
Panels {\it e} and {\it f} show the maximum.  
The shallower dashed lines denote 
$ \vert \mbox{\boldmath$g$} \, - \,
\mbox{\boldmath$g$} _{\rm ex} \vert / \vert 
\mbox{\boldmath$g$} _{\rm ex} \vert \; = \; 
10 ^{-0.5} \, N ^{-1} $ for comparison between panels.
The steeper dashed lines denote
$ \vert \mbox{\boldmath$g$} \, - \,
\mbox{\boldmath$g$} _{\rm ex} \vert / \vert 
\mbox{\boldmath$g$} _{\rm ex} \vert \; = \; 
10 \, N^{-2} $. 

In the cells not adjacent to the grid level boundary, the
simple average and root mean square average of the relative
error in the gravity are proportional to $ N ^{-2} $.
The root mean square average are nearly
as large as the simple average.  
The maximum value is only three times larger than the simple 
average at the levels of $ \ell $ = 1, 2, 3, and 4. 
It is factor 10 larger than the simple average at
$ \ell \, = \, 5 $.  These result confirm that our scheme
gives the second order accurate gravity in the cells
neither adjacent to the grid level boundary nor lying on
the sharp edge of the density distribution.

In the cells adjacent to the grid level boundary, 
the relative error decreases in proportion to $ N ^{-1} $
except at $ \ell \, = \, 1 $.  This is because our
scheme is only the first order accurate at the grid
level boundary.  This does not mean that the error is
always large only at the grid level boundary.
Note that the average relative error is more or less the
same between in the
cells adjacent to the grid level boundary and in the
rest cells when $ N \, \le \, 32 $.
Only when $ N $ is sufficiently large, the error is
appreciably large at the grid level boundary.

At $ \ell \, = \, 1 $, the relative error decreases 
rapidly with increase in $ N $.  When $ N \, \ge \, 128 $,
the binary is resolved even in the grid of $ \ell \, = 1 $.
Then the dipole and quadruple moments of the gravity are
evaluated with a good accuracy and the outer boundary 
condition is improved greatly.  Accordingly the relative
error is small at $ \ell \, = \, 1 $ when $ N \, \ge \, 128 $.
Remember that the error due to the outer boundary condition 
dominates at $ \ell \, = \, 1 $.

\begin{figure}
\epsscale{0.4}
\plotone{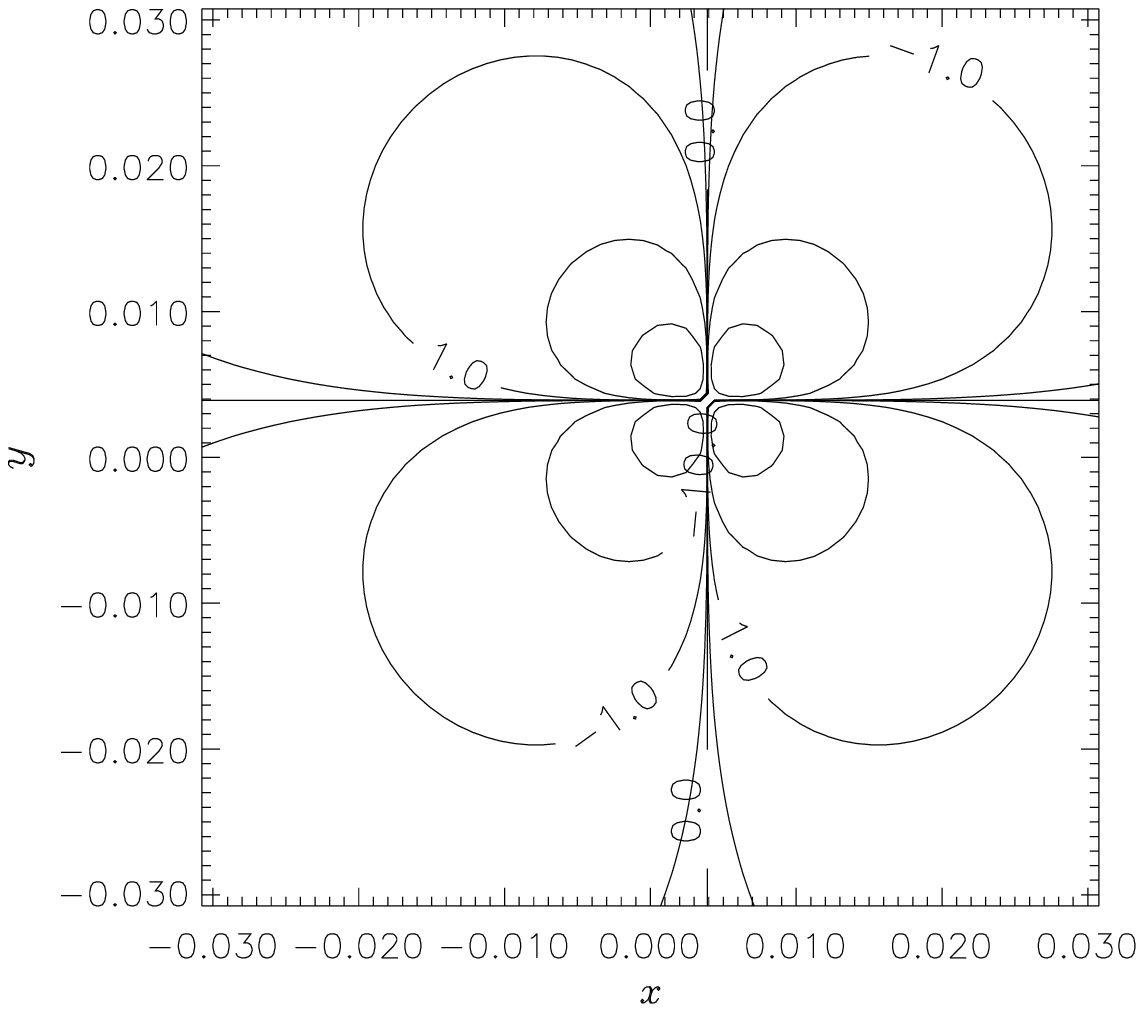}
\plotone{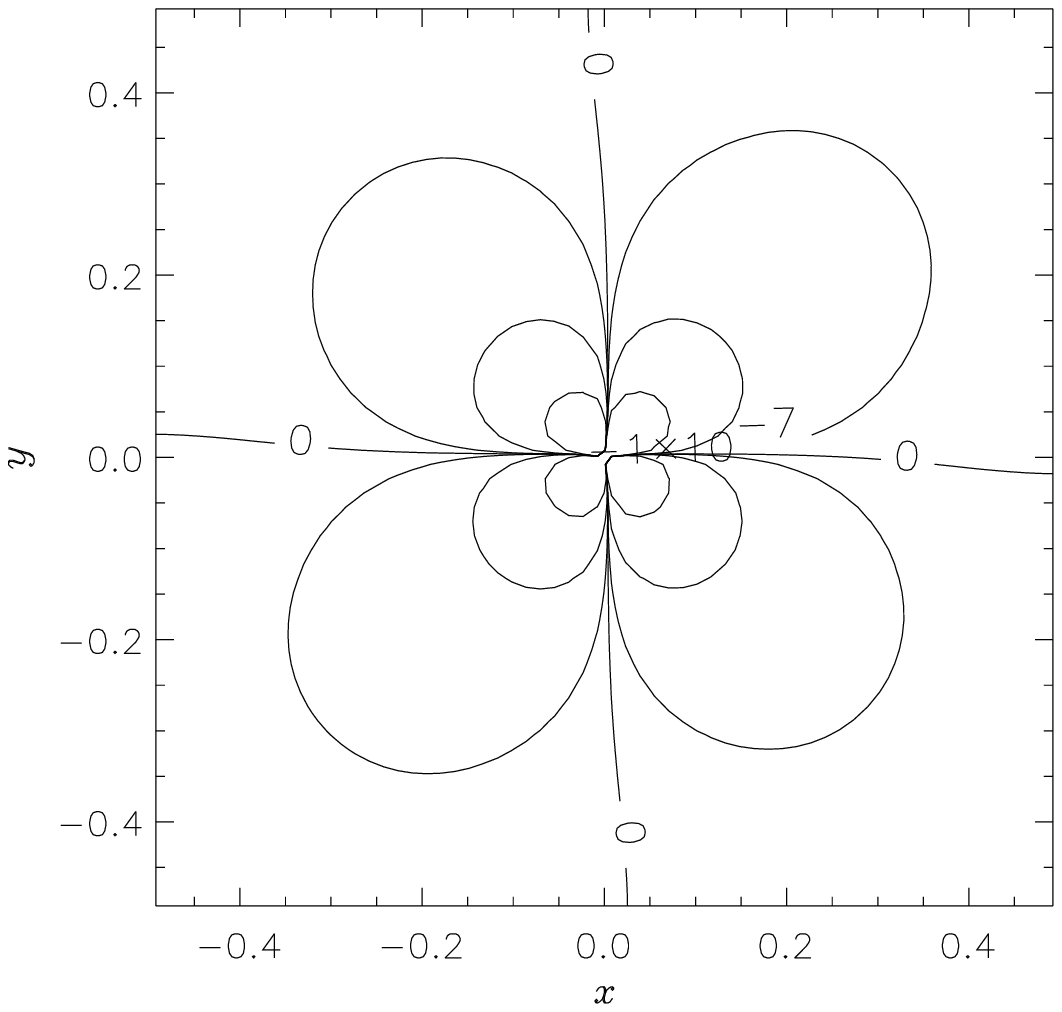}
\caption{Gravitational potential computed for the density
distribution in which four uniform density gas spheres are
placed near the center of the computation grid.  Two of them
have positive mass and the other two have negative mass so that
the gravitational field is well approximated by the quadraple
in the region far from the center.  The location and size of the
four gas spheres are listed in Table 2.
The contours are set in the equal interval in the
logarithmic scale at the level of $ \phi \, = \, \pm 10 ^{n} $
and 0 where $ n $ denotes an arbitrary integer.}
\label{Q-pole.eps}
\end{figure}

We made another example to confirm that our scheme can reproduce
not only the $ 1/r $ gravity but also the quadraple moment of the
gravity accurately.
Figure \ref{Q-pole.eps} shows the second example in which 4 uniform density spheres
are placed near the center of the grid.  The centers, radii, and
masses of the spheres are listed in Table 2.  Note that the total
mass is zero.  When measured in the coarse grid of 
$ \ell \, \le \, 2 $, the cell average density vanishes in all
the cells since $ N \, = \, 64 $ in this example.
In other words, there are no source terms in the 
discrete Poisson equation at the level of $ \ell \, = \, 1 $.
Nevertheless, the quadraple moment of gravity is reproduced in 
Figure \ref{Q-pole.eps}.  This proves that Equation (\ref{gauss}) connects 
the gravitational potential between the coarse and fine grids.

\subsection{Convergence}
\label{convergence}
\begin{figure}
\epsscale{0.7}
\plotone{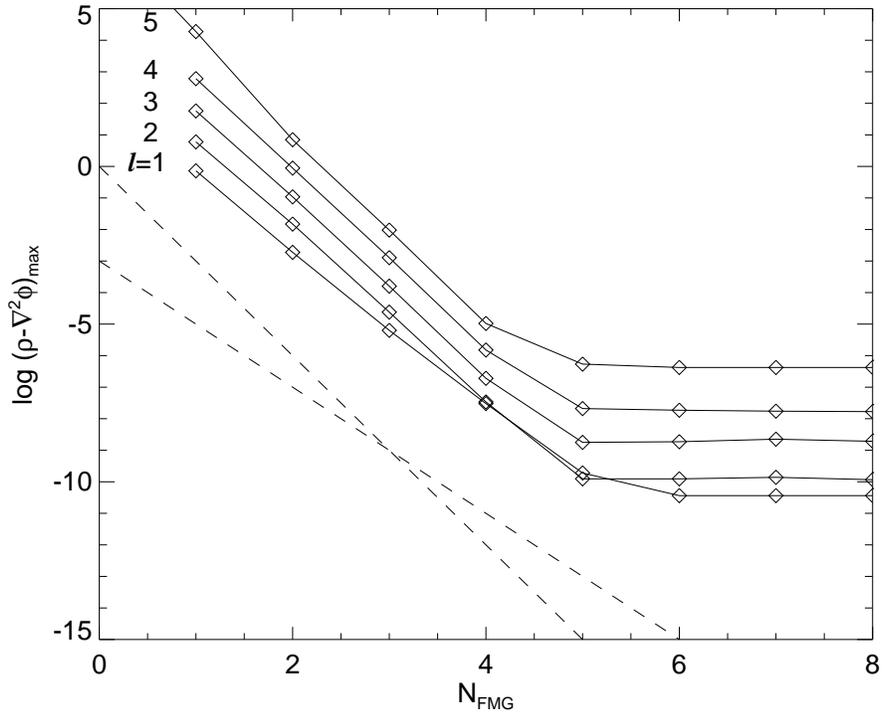}
\caption{ Residual of the Poisson equation ($\rho - \nabla^2\phi$) as
a function of the number of the FMG iteration cycle ($N_{\rm
FMG}$). Each curve denotes the maximum residual measured in the grid of
each level.  The dashed lines denote the relationships of $\rho -
\nabla^2 \phi \propto 10^{-2 N_{\rm FMG}}$ and $10^{-3 N_{\rm FMG}}$,
respectively.  }
\label{res-nmg.eps}
\end{figure}

We measured reduction in the residual, the difference in the right 
and left hand sides of Equation (\ref{dpoisson1}), to evaluate the 
efficiency 
of the iteration. Figure \ref{res-nmg.eps} shows the residual as a function of
the number of the FMG iteration cycle for the sample problem
for binary shown in \S\S \ref{accuracy}.  The absolute value of the residual 
reduces by a factor of 300 per one cycle of multi-grid iteration
in the first four times of the FMG iteration.  The residual
remains constant after the five times FMG iteration.
The remaining residual is due to the round off error in the
gravitational potential.  The residual is proportional
to the inverse square of the cell size in Figure \ref{res-nmg.eps} since
the second derivative is plotted.
The round off error is extremely small and is negligible
compared with the discretization error discussed in \S\S \ref{accuracy}.

Trying several model density distributions, 
we have confirmed that the reduction depends little on the 
spatial distribution of the residual.

When we increase the number of the Gauss-Seidel iteration
during a cycle of the multi-grid iteration, we get a higher 
reduction in the residual per cycle at the expense of 
longer computation time.  The reduction per unit computation
load is higher when we the Gauss-Seidel iteration is 
performed several times each.

We tried the successive over relaxation (SOR) instead the
Gauss-Seidel iteration to accelerate the converge but failed.
When SOR is used as the pre- and post-relaxation in our
multi-grid iteration, the residual increases sometimes
depending on the initial guess.  SOR works well only when 
use one level of the grid, i.e., when the grid is not
nested.

We also tried to improve the interpolation formula,
Equation (\ref{interpolation1}), for a higher order
accuracy.  We found that our iteration did not
converge when we used a higher order interpolation
formula.   Though we can not deny existence of 
a successful interpolation formula, we could not find it.

\subsection{Computation Time}

We evaluated the computation load of our multi-grid algorithm by
measuring the computation time.  A UNIX workstation, SGI O2 (MIPS
R10000 250 MHz) was used for the measurement.  The computation time
was measured with the subroutine DTIME.

\begin{figure}
\epsscale{0.7}
\plotone{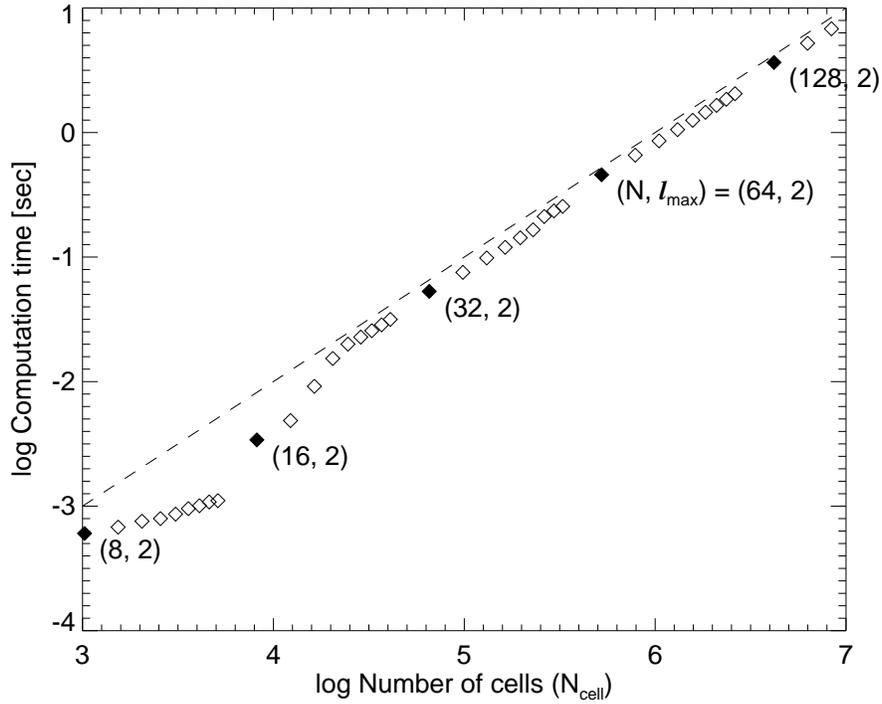}
\caption{Computation time as a function of the number of the cells
in the nested grid.  
The squares denote the computation time measured with
the UNIX workstation, SGI O2 (MIPS R10000 250 MHz).
The ordinate denotes the computation time, $ t $, in unit of second in
the logarithmic scale.
The abscissa denotes the total number of the cells used in the
nested grid, i.e., $ N _{\rm cell} \, = \, {\ell _{\rm max}} \, N ^3 $, 
in the logarithmic scale.  
The parameters of the grid, ($N$, $\ell_{\rm max}$), are labeled
by the filled squares.
The relations of
$ t = 10^{-6} N _{\rm cell} $
are drawn by the dashed lines for comparison.}
\label{etime.eps}
\end{figure}

Figure \ref{etime.eps} shows the execution time per FMG cycle.  When
The abscissa denotes the effective total number, $ N _{\rm cell} $ = $
\ell _{\rm max} \, N ^3 $, in the logarithmic scale.  The ordinate
denotes the logarithm of the computation time, $ t $.  The dashed
lines denote the relations, CPU time $ \propto \, N _{\rm cell} $ and
CPU time.  We made this plot by changing $ \ell _{\rm max} $ and $ N $
in the range of $ 2 \, \le \, \ell \, \le \, 10 $ and $ 8 \, \le \, N
\, \le \, 128 $.

As shown in Figure \ref{etime.eps}, the computation load is proportional
to the $ N _{\rm cell} $ when $ N \, \ge \, 32 $.
This means that our algorithm is scalable for medium and large
nested grids. 

We also measured the computation time with the super computer
Fujitsu VPP5000 at National Astronomical Observatory, Japan.
When $ N \, = \, 256 $ and $ \ell _{\rm max} \, = \, 5 $,
the CPU time is 0.21 sec per FMG cycle.  
This CPU time is reasonably small compared with the 
CPU time for solving the hydrodynamical equation on the
same nested grid.  This CPU time is not scalable to
the number of cells since the
super computer has  vector processors and 
its CPU time is not proportional to the computation load.

\section{DISCUSSIONS}

As shown in the previous section, our numerical method
provides an accurate solution with a reasonably small 
computation cost.  The computation
load is scalable in the sense that it is proportional
to the number of the cells contained in the nested grid.
Our discrete Poisson equation is robust in the
sense that it can be applied also to AMR
as far as a parent cell is subdivided into two in
one dimension.  We discuss the strength of our scheme while
comparing with other schemes given in literatures.

\citet{suisalu95} took another approach when solving
the Poisson equation in their AMR computation.
They adopted an cubic interpolation formula at the
grid level boundary to ensure the continuity of the 
solution.  They have noticed that a linear interpolation ensures 
only the continuity in the potential but not that in the
gravitational force.  Discontinuity in the force may 
cause a spurious feature across the grid level boundary.
Although not mentioned explicitly in \citet{suisalu95},
the discontinuity causes a more  serious problem; it 
introduces spurious mass on the grid level boundary.  
If the gravity $ \mbox{\boldmath$g$} $
has different values at a grid level boundary, 
the solution does not satisfies the Gauss's theorem.
The spurious mass on the grid boundary gives a serious error
on the solution in a long range.  The error decreases only 
inversely proportional to the distance from the spurious mass.
This error dominates over the gravitational torque which decreases
more steeply. 

Compared with adoption of a higher order interpolation formula
at the grid level boundary, our scheme has several advantages.
First our approach ensures absence of a spurious mass on the
grid level boundary while it is not guaranteed in a higher
oder interpolation formula.  Second higher order interpolation
formula increases computation load.  Third higher order 
interpolation may slow down efficiency of convergence. 
As far as in our experience, a simple multi-grid iteration
does not work when Equations (\ref{interpolation1}) and
(\ref{interpolation2}) are replaced with a higher order
interpolation formula.  We discuss the third point further
in the following.

Our discrete Poisson equation has the following favorable
character.  When it is rewritten in the form,
\begin{equation}
\rho _i \; = \; \sum _j
a _{i,j} \phi _j \; ,
\end{equation}
the coefficient $ a _{i,j} $ is always positive except 
for $ i $ = $ j $.  Here symbols $ i $ and $ j $ denote the
cell number after sorted in one dimension.  This property
ensures that the Gauss-Seidel iteration converges always
since it always underestimates the correction.  In other words,
the convergence of the Gauss-Seidel iteration is ensured
since the matrix $ a _{i,j} $ is diagonally dominant,
\begin{equation}
\sum _{i\ne j} \vert a _{i,j} \vert \; \le \;
\vert a _{i,i} \vert \; . \label{Ddominance}
\end{equation}
When we use a higher order interpolation formula, this 
the dominance of the diagonal element [Eq. (\ref{Ddominance})] is lost. 
The coefficient $ a _{i,j} $ is no longer always 
positive.  Then the Gauss-Seidel iteration may overestimate
the correction and may not converge.  If we apply successive
under-relaxation, the convergence slows down.  

\citet{ricker00} have proposed another idea for solving the
Poisson equation in a nonuniform grid.
They have used a nonuniform grid in which the spacing in a 
certain direction varies but only in the direction.
For example, the spacing in the $ x $-direction varies with
$ x $ but not in the $ y $- and $ z $-directions.
Consequently each cell is rectangular and may have different
side lengths.  Each cell has a neighboring cell in each direction
and faces 6 neighboring cells in total except for a cell located
on the grid.  They evaluated the gravity on the cell surface using
the second order interpolation.  Their Poisson equation can be
rewritten as 
\begin{equation}
\frac{g _{x,i+1/2,j,k} \, - \, g _{x,i-1/2,j,k}}
{\Delta x _i} \, + \,
\frac{g _{y,i,j+1/2,k} \, - \, g _{y,i,j-1/2,k}}
{\Delta y _j} \, + \,
\frac{g _{z,i,j,k+1/2} \, - \, g _{z,i-1/2,j,k-1/2}}
{\Delta z _k} \, = \; 4 \pi G \rho _{i,j,k} \; ,
\label{ricker}
\end{equation}
where
\begin{equation}
g _{x,i+1/2,j,k} \; = \; F _{2,i} \phi _{i+2,j,k} \, + \,
F _{1,i} \phi _{i+1,j,k} \, + \, F _{0,i} \phi _{i,j,k} \; 
\end{equation}
\begin{eqnarray}
F _{2,i} & = & 
\frac{2 \, (\Delta x _i \, - \, \Delta x _{i+1})}
{(\Delta x _i \, + \, \Delta x _{i+1} \, + \, \Delta x _{i+2})
\, (\Delta x _{i+1} \, + \, \Delta x _{i+2})} \, \\
F _{1,i} & = & - \,
\frac{2 \, [\Delta x _i ^2 \, - \, 3 \Delta x _{i+1} 
\, ( \Delta x _{i+1} \, + \, \Delta x _{i+2}) \, - \,
\Delta x _{i+2} ^2]}
{(\Delta x _i \, + \, \Delta x _{i+1} \, + \, \Delta x _{i+2})
\, (\Delta x _{i+1} \, + \, \Delta x _{i+2}) \,
(\Delta x _i \, + \, \Delta x _{i+1})} \; , \\
F _{0,i} & = & 
\frac{2 \, (2 \Delta x _{i+1} \, + \, \Delta x _{i+2})}
{(\Delta x _i \, + \, \Delta x _{i+1} \, + \, \Delta x _{i+2})
\, (\Delta x _i \, + \, \Delta x _{i+1})} \; .
\end{eqnarray}
The symbols, $ \Delta x _i $, $ \Delta y _j $, and $ \Delta z _j $
denote the side length of each  cell.
We omitted the interpolation formula
for $ g _y $ and $ g _z $ to save space. Integrating Equation 
(\ref{ricker}) over the whole volume, we obtain 
\begin{equation}
\sum _{i,j,k} \rho _{i,j,k} \, \Delta x _i \, \Delta y _j
\, \Delta z _k \; = \; \sum _{i,j,k} H _{i,j,k} \phi _{i,j,k} \; ,
\end{equation}
where $ H _{i,j,k} $ does not vanish in a nonuniform grid.
This means that $ \phi _{i,j,k} $ is not necessary to vanish even
when $ \rho _{i,j,k} $ = 0.  In other words Equation 
(\ref{ricker}) has multiple solutions for a given density 
distribution and boundary conditions.  This is a serious problem.

As shown above, the approach based on a higher order interpolation
formula has some fundamental problems.   Although our scheme is
only the first order accurate at the grid boundaries, it give
a quantitatively good solution.  This is likely to be due to 
the fact that our difference scheme satisfies global conditions
for the proper Poisson equation to satisfy.  First our scheme
satisfy the Gauss's theorem as mentioned repeatedly.
Second our scheme ensures the Stokes' theorem,
\begin{equation}
\oint \mbox{\boldmath$g$} \cdot d\mbox{\boldmath$s$} \; =
\; 0 \; . \label{stokes}
\end{equation}
Equation (\ref{interpolation1}) is equivalent to
\begin{equation}
\phi (\mbox{\boldmath$r$} _1) \; = \;
\phi (\mbox{\boldmath$r$} _2) \; + \;
\int _1 ^2 \, \mbox{\boldmath$g$} \cdot
d\mbox{\boldmath$r$} \; .
\end{equation}

We adopted a principle that the difference equations 
should be consistent with the global properties, i.e., 
the Gauss's theorem and Stoke's theorem.
We were satisfied with the first order accuracy at the
grid level boundaries.  In other words we gave priority to
the global properties over the higher order accuracy 
at a given point.
As a result we succeeded in computing the gravity of
a close binary and in reproducing the quadraple moment.
This success is analogous to that of a Total Variation
Diminishing (TVD) scheme for wave equations and that 
of symplectic integrator for a Hamiltonian system \citep{yoshida90}.
TVD scheme gives priority to monotonicity of the solution
over the local accuracy \citep[see, e.g.][]{hirsch90}.  
The solution is free from 
numerical oscillations.  A symplectic integrator gives 
a solution which satisfies the conservation of the volume
in phase space.   As a result there is no secular change
in the total energy, even though it is not conserved at each
time step due to the limited accuracy.  
These examples confirm that the global properties are
more important than the higher order accuracy.

An alternative  method was proposed by \citet{truelove98}.
Their numerical scheme is based on that of \citet{almgren98}
who solved the dynamics of an incompressible fluid with AMR.
They evaluated the gravitational potential, $ \phi $, not
on the cell centers but on the vertexes.  While they applied
the cell centered difference to the hydrodynamical
equations, they applied the vertex centered difference
to the Poisson equation.  This method has an advantage that
the different level cells share the same gravitational
potential automatically.  In other words, this method
ensures the continuity of the gravitational potential
between the grids of different levels.  This method, however,
does not satisfies the Gauss's theorem and accordingly 
does not ensure the continuity of the gravity,
$ \mbox{\boldmath$g$} $.  

The method of \citet{truelove98} has another 
disadvantage that it underestimates gravity.  This disadvantage
is due to averaging used to evaluate the vertex centered
density.   They evaluated the density at a vertex since the
Poisson equation is evaluated on the vertex center.
Averaging lowers the peak density and broadens the density 
distribution.  Thus the gravitational potential evaluated
at the vertex is shallower than that evaluated at the
cell center.  This difference is seriously large when 
the gas is concentrated in a few cells.

We have applied our Poisson equation solver to numerical
simulations of binary star formation.   The results will be
published in near future.

\acknowledgments

We thank an anonymous referee for his/her comments
to the earlier version of this manuscript.  They helped
us for improving this manuscript greatly.
We thank the Astronomical Data Analysis Center of the
National Astronomical Society of Japan for allowing us
to use Fujitsu VPP5000.  This study is financially supported
in part by the Grant-in-Aid for Encouragement of Young
Scientists (12740123, 14740134) and that for Scientific Research (C)
(13640237) of Japan Society of Promotion of Science
(JSPS).   It is also supported in part by the Grant-in-Aid
for Scientific Research on Priority Areas (A) (13011204)
of the Ministry of Education, Culture, Sports, Science and
Technology (MEXT).

\newpage

\begin{deluxetable}{cccccc}
\tablecaption{Mass distribution in model shown in Fig. \ref{binary4.eps}}
\tablewidth{0pt}
\startdata
\hline
$ n $ & $ x _{{\rm c},n} $ & $ y _{{\rm c},n} $ & 
$ z _{{\rm c},n} $ & $ r _{{\rm c},n} $ & 
$ M _{\rm n} $ \\
\hline
1 & 12/1024 & 0 & 0 & 6/1024 & 1 \\
2 & $-$12/1024 & 0 & 0 & 6/1024 & 2  
\enddata
\end{deluxetable}

\begin{deluxetable}{cccccc}
\tablecaption{Mass distribution in model shown in Fig. \ref{Q-pole.eps}.}
\tablewidth{0pt}
\startdata
\hline
$ n $ & $ x _{{\rm c},n} $ & $ y _{{\rm c},n} $ & 
$ z _{{\rm c},n} $ & $ r _{{\rm c},n} $ & 
$ M _{\rm n} $ \\
\hline
1 & 2/1024 & 2/1024 & 0 & 2/1024 & 1 \\
2 & 2/1024 & 6/1024 & 0 & 2/1024 & $-1$ \\ 
3 & 6/1024 & 2/1024 & 0 & 2/1024 & $-1$ \\
4 & 6/1024 & 6/1024 & 0 & 2/1024 & 1  
\enddata
\end{deluxetable}

\newpage

\begin{thebibliography}{}
\bibitem[Almgren et al.(1998)]{almgren98} Almgren, A.~S.,
Bell, J.~B., Colella, P., Howell, L.~H., \& Welcome, M.~L.
1998, J. Comput. Phys., 142, 1
\bibitem[Berger \& Oliger(1984)]{berger84}
Berger, M.~J. \& Oliger, J. 1984, J. Comput. Phys., 53, 484  
\bibitem[Burkert \& Bodenheimer(1993)]{burkert93}
Burkert, A., \& Bodenheimer, P. 1993, \mnras, 264, 798
\bibitem[Burkert \& Bodenheimer(1996)]{burkert96}
Burkert, A., \& Bodenheimer, P. 1996, \mnras, 280, 1190
\bibitem[Boss et al.(2000)]{boss00}
Boss, A.~P., Fisher, R.~T., Klein, R.~I., \& McKee, C.~F.
2000, \apj, 528, 325
\bibitem[Briggs, Henson, \& McCormick(2000)]
{briggs00} 
Briggs, W.~L., Henson, V.~E., McCormick, S.~F. 2000,
A Multi Grid Tutorial, Second Ed. (Philadelphia: SIAM)
\bibitem[Hirsch(1990)]{hirsch90}
Hirsh, C. 1990, Numerical Computation of Internal and External Flows,
Vol. 2 (Chichester: Wiley), 528
\bibitem[Ricker et al.(2000)Ricker, Dodelson, \& Lamb]
{ricker00} Ricker, P.~M., Dodelson, S., \& Lamb, D.~Q. 2000,
\apj, 536, 122
\bibitem[Norman \& Bryan(1999)]{norman99} 
Norman, M.~L. \& Bryan, G.~L. 1999, 
in ASSL Vol. 240, Proc. Numerical Astrophysics 
ed. S.~M. Miyama, K. Tomisaka, \& T. Hanawa (Dordrecht: Kluwer) 19
\bibitem[Press et al.(1986)]{press86}
Press, W. H., Flannery, B. P., Teukolsky, S. A., \& Vetterling,
W. T. 1986, Numerical Recipes (Cambridge: Cambridge Univ. Press) 
\bibitem[Press \& Teukolsky(1991)]{press91}
Press, W. H., \& Teukolsky, S. A. 1991, Comput. Phys., 5, 514
\bibitem[Suisalu \& Saar(1995)]{suisalu95} Suisalu, I. \&
Saar, E. 1995, \mnras, 274, 287
\bibitem[Tomisaka(1998)]{tomisaka98} Tomisaka, K. 1998, \apjl,
402, L163
\bibitem[Truelove et al.(1997)]{truelove97}
Truelove, J. K., Klein, R. I., Mckee, C. F., Holliman, J. H., II,
Howell, L. H., \& Greenough, J. A. 1997, \apjl, 489, L179
\bibitem[Truelove et al.(1998)]{truelove98} Truelove, J.~K.,
Klein, R.~I., McKee, C.~F., Holliman, J.~H., II,
Howell, L.~H., Greenough, J.~A., \& Woods, D.~T. 1988,
\apj, 495, 821
\bibitem[Wesseling(1992)]{wesseling92} Wesseling, P. 1992,
An Introduction to Multigrid Methods (Chichester: Wiley)
\bibitem[Yoshida(1990)]{yoshida90} Yoshida, H. 1990, Physics Letters A,
150, 262
\end{thebibliography}
\end{document}